\documentclass[PAPER, cernpreprint, atlasdraft=false,  UKenglish, texmf, orcidlogo]{atlasdoc}
\usepackage[subfigure]{atlaspackage}
\usepackage{atlasbiblatex}

\usepackage{atlasphysics}
\graphicspath{{logos/}{figures/}}

\usepackage{ANA-FTAG-2024-07-PAPER-defs}

\hypersetup{pdftitle={ATLAS document},pdfauthor={The ATLAS Collaboration}}

\AtlasTitle{Simultaneous efficiency measurements of $b$- and $c$-jets in $t\bar{t}$ events
from $\sqrt{s}=13.6$~TeV $pp$ collision data collected with the ATLAS detector}

\AtlasAbstract{%
This paper reports simultaneous measurements of
the efficiencies with which $b$-jets are identified, and $c$-jets are
misidentified as $b$-jets, by the single-transformer-based GN2 tagger that
is currently used in the ATLAS experiment at the LHC.
The measurements use $t\bar{t}$ events where one of the
$W$ bosons decays into an electron or muon and a neutrino, and the
other decays into a quark--antiquark pair, together with the known
$t\to bW$ and $W\to cs$ branching fractions.
The data were collected from $\sqrt{s} = 13.6$~TeV proton--proton collisions
by the ATLAS detector and correspond to an integrated luminosity of 56~fb$^{-1}$.
Events are reconstructed using a kinematic likelihood technique which maps the
jets to the $t\bar{t}$ decay products.
The $b$-jet tagging efficiency is measured from jets arising from leptonically decaying
top quarks and the $c$-jet mistagging efficiency is obtained from jets arising
from hadronic $W$ decays.
The efficiencies are measured as a function of jet transverse momentum
in intervals of the tagging-variable distribution. The $b$-jet ($c$-jet) tagging efficiency measurements reach a precision of better than 2\% (5\%).
Each $b$-jet efficiency is combined with an independent measurement
using $t\bar{t}$ events where both $W$ bosons decay leptonically.

}

\AtlasRefCode{FTAG-2024-07}

\PreprintIdNumber{CERN-EP-2025-292}

\AtlasJournal{Eur.\ Phys.\ J.\ C}
\AtlasCoverEgroupAnalysisTeam{atlas-FTAG-2024-07-analysis-team@cern.ch}

\begin{document}

\maketitle

\tableofcontents

%
%

%
%

\section{Introduction}
\label{sec:intro}
In many analyses at the Large Hadron Collider (LHC)~\cite{Evans:2008zzb}
it is crucial to identify which jets originate from a bottom or charm quark;
recent examples using data collected by the
ATLAS experiment~\cite{PERF-2007-01} are a measurement of Higgs bosons
decaying to bottom quarks and search for Higgs bosons decaying to charm
quarks~\cite{HIGG-2020-20}, measurements of top-quark pair
production with additional \bjets~\cite{TOPQ-2019-03} or
\cjets~\cite{TOPQ-2021-26}, and searches for Higgs boson pair
production~\cite{HIGP-2025-10} and Higgs boson production in association
with top quarks~\cite{HIGG-2020-24}.

The algorithms which identify jets that originate from a bottom quark
are known as \btagging algorithms. The performance of these algorithms
depends crucially on whether the jets that are
identified contain $b$-hadrons (\bjets), $c$-hadrons (\cjets), or neither
of them (light-flavour jets).
Monte Carlo (MC) simulation provides an estimate of the probability that
a jet fulfils the requirements of the \btagging algorithm, i.e.\ the
tagging efficiency  for \bjets\ and the mistagging
efficiency\footnote{Mistagging efficiency is the efficiency for an object other than a $b$-jet to be identified as a $b$-jet by the \btagging algorithm.}
for \cjets, light-flavour jets or jets formed from
the products of hadronically decaying $\tau$-leptons. However, since the
simulation is not perfect, the (mis)tagging efficiency is
measured in data. Simulation-to-data scale factors (SFs), which are defined as the ratio of
the efficiency measured in data to that in simulation, are used to
correct the simulation, assuming that the SFs
are independent of the physics process. Analyses that use
\btagging  apply a weight to each simulated event, calculated as the
product of the SFs for all jets in the event.

ATLAS recently measured the SFs for \bjets~\cite{FTAG-2018-01} in data
proton-proton ($pp$) data at a centre-of-mass energy of 13~TeV from 2015-2018  (LHC \RunTwo). That measurement used dileptonic top-quark pair production (\ttbar)
where the $W$ bosons from both top-quark decays produce an electron
or a muon plus the corresponding neutrino.
Measurements of SFs for \cjets\ used a method similar to that presented in
this paper, utilising \ttbar\  semileptonic  events where the $W$ boson
originating from one of the top quarks decays leptonically into an electron or
a muon plus the corresponding neutrino ($t\to Wb \to \ell \nu b$) and the
other decays hadronically ($t\to Wb \to qq^\prime b$)~\cite{FTAG-2020-08}.
The SFs for the mistagging efficiency for light-flavour jets were measured
in $Z$+jet events~\cite{FTAG-2019-02}.

This paper is the first to report simultaneous measurements of the \bjet\ and \cjet\
SFs using \ttbar\ semileptonic events.  The measurements use data from
13.6~TeV \(pp\) collisions recorded with the ATLAS detector during
part of \RunThr in 2022 and 2023, corresponding to an
integrated luminosity of \lumi~\cite{ATL-DAPR-PUB-2024-001,LUCID2}.
The method for extracting the \cjet\ SFs was updated for use with
an improved transformer-based graph neural-network flavour tagger,
since this tagger's lower \cjet\ mistagging efficiency~\cite{FTAG-2023-05}
makes the measurements of these SFs more challenging.
A likelihood-based method is used to select a high-purity semileptonic \ttbar\ sample.
The leptonically decaying top quark is used to determine the \bjet\ SFs.
The known branching fraction of the $W$ boson to a final state with a $c$-quark,
approximately 33\%~\cite{Zyla:2020zbs}, allows the \cjet\ mistagging efficiency
to be determined from the data by using the $W$ boson from the hadronically decaying top quark.
The SFs are determined as a function of jet transverse momentum (\pt),
as is also the case in Ref.~\cite{FTAG-2020-08}.
%
%
%

%
%
%
%
%

%
%
%
%
%
%
%
%
%


%
%
%

%
\newcommand{\AtlasCoordFootnote}{%
ATLAS uses a right-handed coordinate system with its origin at the nominal interaction point (IP)
in the centre of the detector and the \(z\)-axis along the beam pipe.
The \(x\)-axis points from the IP to the centre of the LHC ring,
and the \(y\)-axis points upwards.
Polar coordinates \((r,\phi)\) are used in the transverse plane,
\(\phi\) being the azimuthal angle around the \(z\)-axis.
The pseudorapidity is defined in terms of the polar angle \(\theta\) as \(\eta = -\ln \tan(\theta/2)\) and is equal to the rapidity
$ y = (1/2) \ln[(E + p_z)/(E - p_z)] $ in the relativistic limit.
Angular distance is measured in units of \(\Delta R \equiv \sqrt{(\Delta y)^{2} + (\Delta\phi)^{2}}\).}

\section{ATLAS detector}
\label{sec:detector}
The ATLAS experiment~\cite{PERF-2007-01,GENR-2019-02} at the LHC uses a multipurpose particle detector
with a forward--backward symmetric cylindrical geometry and nearly \(4\pi\) coverage in
solid angle.\footnote{\AtlasCoordFootnote}
It consists of an inner tracking detector surrounded by a thin superconducting solenoid
providing a \qty{2}{\tesla} axial magnetic field, electromagnetic and hadronic calorimeters, and a muon spectrometer.
The inner tracking detector covers the pseudorapidity range \(|\eta| < 2.5\).
It consists of silicon pixel, silicon microstrip, and transition radiation tracking detectors.
Lead/liquid-argon (LAr) sampling calorimeters provide electromagnetic (EM) energy measurements
with high granularity within the region \(|\eta|< 3.2\).
A steel/scintillator-tile hadronic calorimeter covers the central pseudorapidity range (\(|\eta| < 1.7\)).
The endcap and forward regions are instrumented with LAr calorimeters
for EM and hadronic energy measurements up to \(|\eta| = 4.9\).
The muon spectrometer surrounds the calorimeters and is based on
three large superconducting air-core toroidal magnets with eight coils each.
The field integral of the toroids ranges between \num{2.0} and \qty{6.0}{\tesla\metre}
across most of the detector.
The muon spectrometer includes a system of precision tracking chambers up to \(|\eta| = 2.7\) and fast detectors for triggering up to \(|\eta| = 2.4\).
The luminosity is measured mainly by the LUCID-2 detector, which is located close to the beam pipe.
A two-level trigger system is used to select events~\cite{TRIG-2022-01}.
The first-level trigger is implemented in hardware and uses a subset of the detector information
to accept events at a rate below \qty{100}{\kHz}.
This is followed by a software-based trigger that
reduces the accepted event rate to \qty{3}{\kHz} on average,
depending on the data-taking conditions.
A software suite~\cite{SOFT-2022-02} is used in data simulation, in the reconstruction
and analysis of real and simulated data, in detector operations, and in the trigger and data acquisition
systems of the experiment.

%
%
%
%
%
%
%
%
%
%
%
%
%
%
%
%
%
%
%
%
%
%
%
%
%
%
%
%
%
%
%
%
%
%
%
%
%
%
%
%
%
%
%
%
%
%
%
%
%
%
%
%
%
%
%
%
%
%
%
%
%
%
%
%
%
%
%
%
%
%
%
%
%
%
%
%
%
%
%
%
%
%
%
%
%
%
%
%
%
%
%
%
%
%
%
%
%
%
%
%
%
%
%
%
%
%
%
%
%
%
%
%
%
%
%
%
%
%
%
%
%
%
%
%
%
%
%
%
%
%
%
%


%
\section{Data and simulated event samples}
\label{sec:samples}
The data analysed in this paper correspond to
\lumi\  of
\(pp\) collision data collected by the ATLAS detector in 2022 and
2023 with a centre-of-mass energy of 13.6~\TeV\ and a 25\,ns proton
bunch crossing interval.  The data sample was collected using a set of
single-electron and single-muon triggers  with lowest \pt\ thresholds
of 26~\GeV for electrons and 24~\GeV for
muons~\cite{TRIG-2022-01,TRIG-2018-05,TRIG-2018-01}.
All detector subsystems were required to be operational
during data-taking~\cite{DAPR-2018-01}. Events with noise bursts or coherent
noise in the calorimeters are removed.  The
presence of additional interactions in the same or a neighbouring bunch crossing,
referred to as pile-up, is characterised by the average number of such
interactions, \averagemu, which was 42 (51) during 2022 (2023).

Simulated event samples are used to model Standard Model
processes and to estimate the expected signal yields.  All samples were
produced using the ATLAS simulation infrastructure~\cite{SOFT-2010-01}
and $\GEANT$~\cite{Agostinelli:2002hh}.
The simulated events are reconstructed with the same algorithms as
used for data, and contain a realistic modelling of pile-up
interactions.
The pile-up profiles in the simulation match those of the data
and were obtained by overlaying the hard-scatter events with
minimum-bias events simulated using the soft QCD processes
generated from a mix of \EPOS[2.0.1.4]~\cite{Werner:2005jf}
and \PYTHIA[8.308]~\cite{Bierlich:2022pfr}. The \EPOS events
were generated with parameter values from the \EPOS LHC tune~\cite{Pierog:2013ria}
and the \PYTHIA events with the A3 tune~\cite{ATL-PHYS-PUB-2016-017}
and the \NNPDF[2.3lo]~\cite{Ball:2012cx} set of parton distribution
functions (PDFs). \PYTHIA pile-up events include either a high-$\pt$ jet,
a prompt photon, or a lepton from a $b$-hadron decay, while \EPOS was
filtered to simulate all remaining pile-up events in the overlay sample.
For all samples, with the exception of those generated using
\SHERPA~\cite{Bothmann:2019yzt}, the decays of bottom and charm
hadrons were performed by \EVTGEN~\cite{Lange:2001uf}.

The events used in this study originate mostly from \ttbar\
production. This process was modelled using the
\POWHEGBOX[v2]~\cite{Frixione:2007nw,Nason:2004rx,Frixione:2007vw,Alioli:2010xd}
generator at next-to-leading order (NLO) with the \NNPDF[3.0nlo] PDF
set~\cite{Ball:2014uwa} and the \hdamp\ parameter\footnote{The \hdamp\
parameter is a resummation damping factor and one of the parameters
that control the matching of \POWHEG matrix elements to the parton
shower and thus effectively regulate the high-\pt\ radiation against
which the \ttbar\ system recoils.} set to $1.5 \times
\mtop$~\cite{ATL-PHYS-PUB-2016-020}, where \mtop\ denotes the mass of
the top quark.  The events were interfaced to \PYTHIA[8.308] to model
the parton shower, hadronisation, and underlying event, with
parameters set according to the A14 tune~\cite{ATL-PHYS-PUB-2017-007}
and using the \NNPDF[2.3lo] set of PDFs~\cite{Ball:2014uwa}. The
production was split into three parts: semi-leptonic decays, dileptonic
decay and a high statisics sample for events with the decay to $W\to c
b$, which was produced for the measurement of the CKM
element $V_{cb}$~\cite{TOPQ-2019-14}.

The uncertainty due to initial-state radiation (ISR) is estimated by
varying the \hdamp\ parameter and the renormalisation and
factorisation scales, and choosing the Var3c up/down variants of the
A14 tune as described in Ref.~\cite{ATL-PHYS-PUB-2017-007}.
The impact of final-state radiation (FSR) is evaluated by varying the
renormalisation scale for emissions from the parton shower by a factor two.
The impact of using a different parton shower and
hadronisation model is evaluated by comparing the nominal \ttbar\ sample
with another \ttbar\ sample generated by \POWHEGBOX[v2]
using exactly the same
parameter values but interfaced with \HERWIG[7.13]~\cite{Bahr:2008pv,Bellm:2015jjp},
using the default tune~\cite{Bellm:2015jjp,Bellm:2017jjp}
and the \MMHT[lo] PDF set~\cite{Harland-Lang:2014zoa}.
Finally, the matrix element matching uncertainty was evaluated
by setting the \pthard\  parameter to 1 instead of its nominal value of
0~\cite{ATL-PHYS-PUB-2016-020}.

In addition to \ttbar\ production, some minor backgrounds
contribute to the final event sample used for the calibration.
These backgrounds consist mostly of single-top and diboson production,
the production of \ttbar\ in association with a vector boson
and the production of a vector boson in association with jets.
Details of the modelling of these samples are given in the following.

Single-top $s$-channel production was modelled using the \POWHEGBOX[v2]
generator~\cite{Alioli:2009je} at NLO in QCD in the five-flavour scheme with the \NNPDF[3.0nlo] PDF set.
Single-top $t$-channel production was modelled using the
\POWHEGBOX[v2]
generator~\cite{Alioli:2009je} at NLO in QCD, using the four-flavour scheme and the
corresponding \NNPDF[3.0nlo] set of PDFs.
The associated production of a single top quark and a $W$ boson ($tW$) was
modelled using the
\POWHEGBOX[v2]
generator~\cite{Re:2010bp} at NLO in QCD, using the five-flavour scheme and the
\NNPDF[3.0nlo] set of PDFs.
The diagram removal scheme~\cite{Frixione:2008yi} was used to
remove interference and overlap with \ttbar\ production.
The events for all single-top production channels
were interfaced to \PYTHIA[8.308], which
used the A14 tune
and the \NNPDF[2.3lo] set of PDFs.

The production of $Z$+jets and $W$+jets events was simulated with the
\SHERPA[2.2.14]
generator using NLO matrix elements for up to two partons, and leading-order (LO) matrix elements
for up to five partons calculated with the Comix~\cite{Gleisberg:2008fv}
and \OPENLOOPS~\cite{Buccioni:2019sur,Cascioli:2011va,Denner:2016kdg} libraries. They
were matched with the \SHERPA parton shower~\cite{Schumann:2007mg} using the MEPS@NLO
prescription~\cite{Hoeche:2011fd,Hoeche:2012yf,Catani:2001cc,Hoeche:2009rj}
and the set of tuned parameters developed by the \SHERPA authors.
The \NNPDF[3.0nnlo] set of PDFs~\cite{Ball:2014uwa} was used and their predictions
were normalised to a next-to-next-to-leading-order (NNLO)
prediction~\cite{Anastasiou:2003ds}.

Samples of events with diboson final states ($VV$) were simulated with the
\SHERPA[2.2.14]
generator,
including off-shell effects and Higgs boson contributions where appropriate.
Fully leptonic final states and semileptonic final states, where one boson
decays leptonically and the other hadronically, were generated using
matrix elements at NLO accuracy in QCD for up to one additional parton
and at LO accuracy for up to three additional parton
emissions.
The matrix element calculations were matched
and merged with the \SHERPA parton shower based on Catani--Seymour
dipole factorisation
using the MEPS@NLO
prescription.
The virtual QCD corrections were provided by the
\OPENLOOPS library. The
\NNPDF[3.0nnlo] set of PDFs was used,
along with the dedicated set of tuned parton-shower parameters developed by the
\SHERPA authors.

Production of \ttbar\ in association with a vector boson
was modelled using the
\MGNLO[3.5.1] \cite{Alwall:2014hca} generator at NLO with the
\NNPDF[3.0nlo] PDF set.
The events were interfaced to \PYTHIA[8.309]~\cite{Sjostrand:2014zea}
using the A14 tune and the \NNPDF[2.3lo] PDF set.


%


%
\section{Object selection}
\label{sec:objects}

Selected events are required to contain at least one vertex having at least two associated tracks with $\pT > 500$~\MeV, and the primary vertex is chosen to be the reconstructed vertex with the largest $\Sigma p_{\mathrm{T}}^2$ of its associated tracks.

Electron candidates are reconstructed by matching inner-detector
tracks to clusters of energy deposited in the EM calorimeter.
Electrons must have $\pt>27$~\GeV\ and $|\eta|<2.47$. The
associated track must have $|d_0|/\sigma_{d_0}<5$ and $|z_0|\sin
\theta<0.5$~mm, where $d_0$ ($z_0$) is the transverse (longitudinal)
impact parameter relative to the primary vertex, $\sigma_{d_0}$ is
the uncertainty in $d_0$, and $\theta$ is the polar angle of the track.
Candidates are identified with a likelihood method and must satisfy the `medium' identification
criteria described in Ref.~\cite{EGAM-2018-01}. The likelihood relies
on the shape of the EM shower measured in the calorimeter, the quality
of the track reconstruction, and the quality of the match between the
track and the cluster. To suppress electrons originating from photon
conversions, hadron decays, or jets misidentified as electrons,
candidates are required to satisfy the Tight\_VarRad isolation criteria
defined in Ref.~\cite{EGAM-2021-01}. These criteria are based on constraints on the energy deposited in the calorimeter and the sum of the track transverse momenta in a variable-size cone around the electron candidate.
The electron energy and reconstruction, identification and trigger efficiencies are calibrated as described in Refs.~\cite{EGAM-2021-01, EGAM-2021-02}.

Muon candidates are reconstructed in the range $|\eta|<2.5$ by
combining tracks in the inner detector with tracks in the muon
spectrometer.
All muon candidates must have $\pt>27$~\GeV,
$|d_0|/\sigma_{d_0}<3$, and $|z_0|\sin \theta<0.5$~mm.  The `medium' quality
requirements  are used
and muons from hadron decays are suppressed by imposing the `tight' isolation
requirement with a variable-size cone as described in Ref.~\cite{MUON-2018-03}.
The muon momentum is calibrated as described in Ref.~\cite{MUON-2022-01}, and the reconstruction efficiency in the
simulation is corrected to agree with that measured for muons from $Z\to\mu\mu$ decays in data~\cite{MUON-2018-03}.

Jets are formed using objects from a
particle-flow  algorithm, which combines  energy
deposits in the calorimeter with inner detector
tracks~\cite{PERF-2015-09}. The particle-flow objects are combined into
jets  using the anti-$k_t$
algorithm~\cite{Cacciari:2008gp,Fastjet} with a radius parameter $R$ of
$0.4$. Only jets within the range in the range $|\eta|<2.5$ and $\pt>20$~\GeV\ are used for the results presented in this paper. A jet-vertex-tagging
technique using a multivariate likelihood~\cite{PERF-2014-03, ATL-PHYS-PUB-2019-026} is
applied to jets with $\pt<60$~\GeV\ to suppress jets
that are not associated with the event's primary vertex.
Jets are
further calibrated according to in situ measurements of the jet energy
scale~\cite{JETM-2018-05}.

The labelling
scheme used to define the flavour of a jet in a simulated event is applied by
matching reconstructed jets to  generator-level $b$- or $c$-hadrons
with $\pt > 5$~\GeV\ within a cone of size $\Delta R = 0.3$ around the
jet axis. Jets that contain a $b$-hadron are called \bjets.
Remaining jets containing a $c$-hadron are called \cjets. Jets with a
hadronically decaying $\tau$-lepton are called $\tau$-jets, and all remaining
jets are called light-flavour jets.

The larger mass and longer lifetime of the $b$-hadron are used to
distinguished \bjets\ from other jets .  The \btagging algorithm
studied in this paper is called GN2 and is described in detail in
Ref.~\cite{FTAG-2023-05}.  GN2 differs from previous taggers used in
the ATLAS experiment in that track and jet information is used
directly in the training, without running dedicated secondary-vertex
reconstruction algorithms. The GN2 tagger uses a single-transformer
model~\cite{vaswani2023attentionneed} with a single discriminant as
output.  The GN2 tagger is used in discrete intervals defined by the
efficiency to tag \bjets\ in simulated $t\bar{t}$
events~\cite{FTAG-2018-01}: 0--65\%, 65--70\%, 70--77\%, 77--85\%, 85--90\%
and 90--100\%. The 90--100\% interval is  where most of the light-flavour jets are found and is referred to as the untagged bin. Jets with GN2 values that
do not lie in this interval are described as tagged.

Significant differences are seen between the simulated (mis)tagging efficiency for different
fragmentation models~\cite{FTAG-2018-01}. To account for these
differences, simulation-to-simulation SFs are applied to
those simulated samples that have a fragmentation model and decay different
from the one used to
derive the SFs. In the present paper the
$t\bar{t}$ simulation (see Section~\ref{sec:samples}) uses
\PYTHIA[8] and \EVTGEN, which is the reference MC fragmentation program for all the measured SFs.

Overlaps between reconstructed objects are removed using a procedure
based on the angular separation between different final-state objects.
The procedure is the same as the one described in Ref.~\cite{TOPQ-2018-15}.

The event's missing transverse momentum \METvec, whose magnitude is denoted by
\MET, is computed as the negative vectorial sum of the transverse
momenta of leptons,
jets
and a track-based soft term~\cite{JETM-2020-03} accounting for the
contribution from particles from the primary vertex that are not
already included. The jets employed in the \METvec\ calculation
include particle-flow jets and, in addition, jets  with $\pt>30$~\GeV\ and
$2.5<|\eta|<4.5$, which are formed solely with calorimeter energy clusters.


%

\section{Event selection}
\label{sec:eventsel}
The analysis aims to obtain a pure sample of semileptonic \ttbar\
events.  Exactly one electron or one muon, denoted by $\ell$, is
required.  The
events are also required to have $\MET > 20$~\GeV, and the transverse
mass $\mt$ of the lepton  and \METvec\ must satisfy
\[ \mt = \sqrt{2 \pt^\ell \MET (1-\cos\Delta\phi)} > 40~\GeV,\] where
$\Delta\phi = \phi(\METvec)-\phi(\ell)$ is the azimuthal angle between
the lepton and \METvec.

In order to accept events with a high number of jets covering a
large range in $\pt$, while reducing the non-$t\bar{t}$ background and
the number of jets arising from initial- and final-state QCD radiation (ISR and FSR),
requirements are made such that either of the following selections is true:
\begin{itemize}
\item the event contains exactly four jets with $\pt > 25$~\GeV\
\item the event contains at least three jets with $\pt > 25$~\GeV\
and exactly one jet with $20 < \pt < 25$~\GeV
\end{itemize}
The first selection accepts ISR/FSR jets at a relatively
low rate because it requires exactly four jets with a moderate $\pt$ threshold.  The second selection allows
measurements of events with a jet with \pt\  of 20--25~\GeV, while minimising
the non-$t\bar{t}$ background, which is more likely to have multiple low-$\pt$ jets.
In order to improve the statistics at higher $\pt$ in the \cjet\ SF determination, events with more than four jets are allowed if

\begin{itemize}

\item the event contains at least five jets with $\pt > 25$~\GeV\
and at least one of these jets has  $\pt > 70$~\GeV.
\end{itemize}

The  \METvec\ and the four-momenta of the four highest-\pt\ jets and the lepton are used as inputs to a likelihood-based \ttbar\ event
reconstruction algorithm, which minimises deviations of the invariant
masses of the top quarks and $W$ bosons from their true values
and is described in more detail in
Ref.~\cite{Erdmann:2013rxa}. This algorithm uses a likelihood function
to assign the four jets to the \ttbar\ decay topology. In particular,
the algorithm assigns one jet to the \bjet\ from the leptonically
decaying top quark,  another jet to  the \bjet\
from the hadronically decaying top quark and
the remaining two jets to the jets that come from the hadronic $W$
boson decay.
The jet assignment does not use any \btagging information.
The following notation is used: the jets that are
assigned as the decay products of the $W$ boson are referred to as
$W$-jets and the remaining two jets are referred to as the leptonic top-jet and hadronic top-jet.

Among all the possible four-jet permutations, the one with the highest likelihood is selected.
The logarithm of the likelihood is required to be greater than
$-48$~\cite{FTAG-2020-08} so as to increase the fraction
of events where the top-quark decay products are correctly assigned.
%
%
%
%
%
%

%
%
%
%
%
%
%
%
%
%
%
%
%
%
%
%
%
%
%
%
%
%
%
%
%
%
%
%
%
%

%
%
%
%
%
%
%

%
%
%
%
%
%
%
%
%
%
%
%
%
%
%
%
%


%
%

%

\section{Simultaneous efficiency measurements of $b$- and $c$-jets in \ttbar\
events}
\label{sec:analysis}
\subsection{Analysis strategy}
\label{sec:method}

\begin{figure}%
\centering
\subfigure[Pre-fit]{\includegraphics[width=0.45\textwidth]{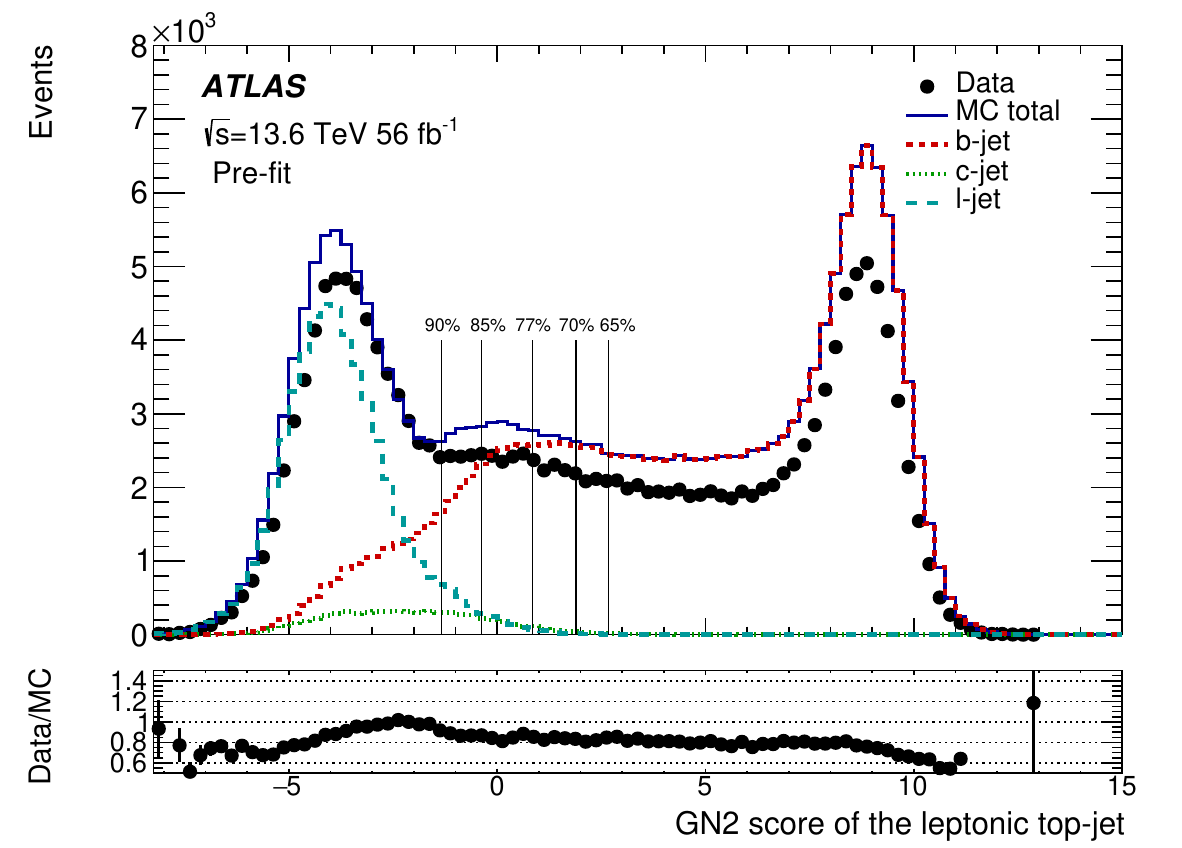} \label{fig:GNprefit} }
\subfigure[Post-fit]{\includegraphics[width=0.45\textwidth]{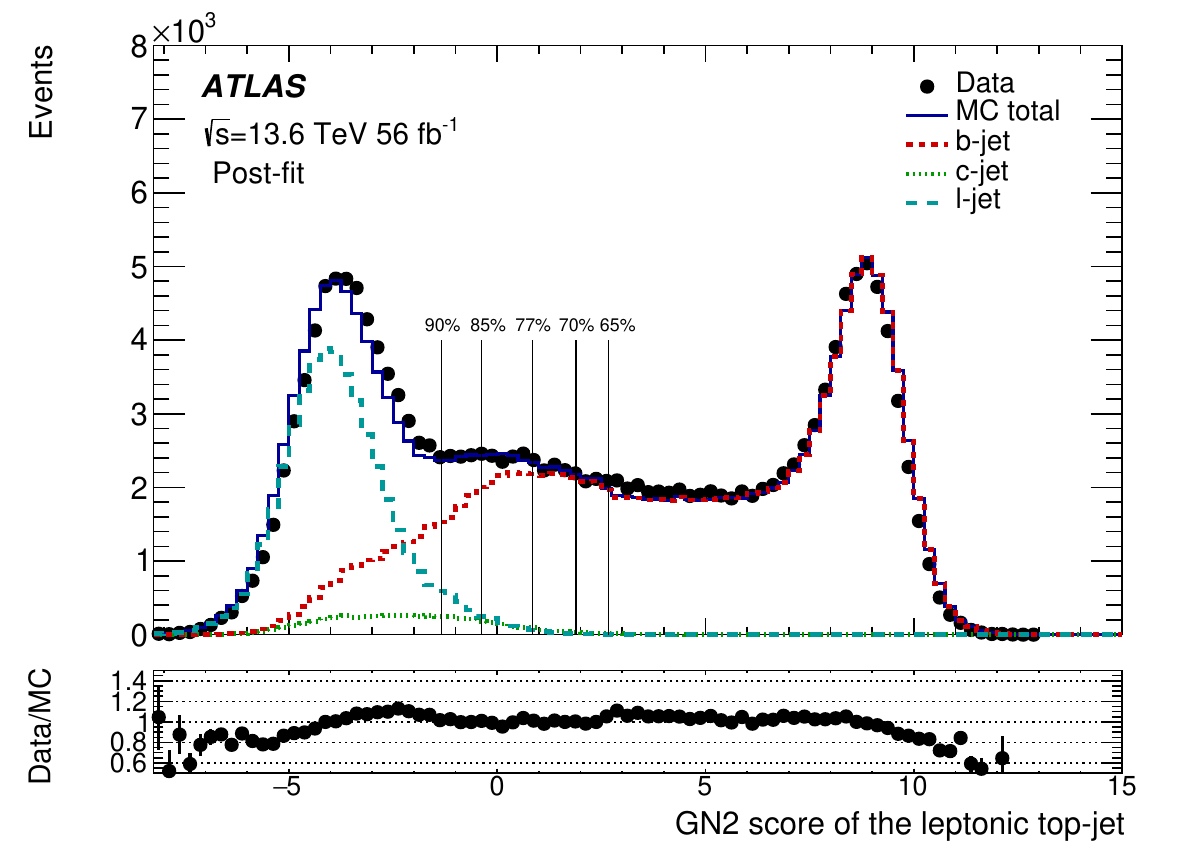} \label{fig:GNpostfit} }
\caption{Distributions of the GN2 discriminant of the leptonic top-jet for events where the hadronic top-jet is tagged in the $0{-}65\%$ interval
and both $W$-jets are untagged.  The  simulation
is split according to the flavour of the jet. The normalisations and \bjet\ tagging and \cjet\ mistagging
efficiency scale factors, determined in the fit described in this section, are not applied to the simulation in
(a) but are applied in (b). The distributions are shown
without systematic errors.
The dashed vertical lines demarcate the tagging intervals in
which the measurements of the (mis)tagging efficiency scale factors
are made.
}
\label{fig:GN}
\end{figure}

The \bjet\ efficiency SFs $b^{t}(i)$ and \cjet\ mistagging efficiency
SFs $c^{t}(i)$ are determined in jet-\pt\ intervals $i$, with
boundaries at 20, 40, 65, 140 and 250~\GeV, and tagging intervals $t$,
where $t=0$ is the untagged bin ($90{-}100\%$ interval), $t=1$ is the
$85{-}90\%$ interval and so on until $t=5$, which corresponds to the
$0{-}65\%$ interval. In all events the hadronic top-jet is required to be
tagged in the tightest ($0{-}65\%$) interval, in order to reduce both the rate
of top-quark decays with wrongly assigned jets and the non-$t\bar{t}$ background. The light-flavour jet SFs
$l^{t}(i)$ are determined independently of the analysis presented
here, but using the same dataset and tagger, using jets produced in
association with a $Z$ boson and a method similar to that described
in Ref.~\cite{FTAG-2019-02}; they are applied in all distributions
shown in this paper.  The method described below is for the
`pseudo-continuous' calibration where the efficiency is defined as the
fraction of jets that have a GN2 discriminant value that lies between the
boundaries of each tagging interval, i.e.\ above the 65\% boundary (as
illustrated in Figure~\ref{fig:GN}), between the 70\% and 65\% boundaries, etc.
Two different event selections targeting $b^{t}(i)$ and $c^{t}(i)$ are employed,
with some events common to both, as described below. All parameters of interest
are determined in a single fit  by minimising a total $\chi^2$ defined as
\begin{equation}
\chi^2 = \chi_b^2 +  \chi_c^2 +\chi_\epsilon^2,
\label{eqn:chi2}
\end{equation}
with the terms defined below.

\begin{figure}%
\centering
\subfigure[Pre-fit]{\includegraphics[width=0.45\textwidth]{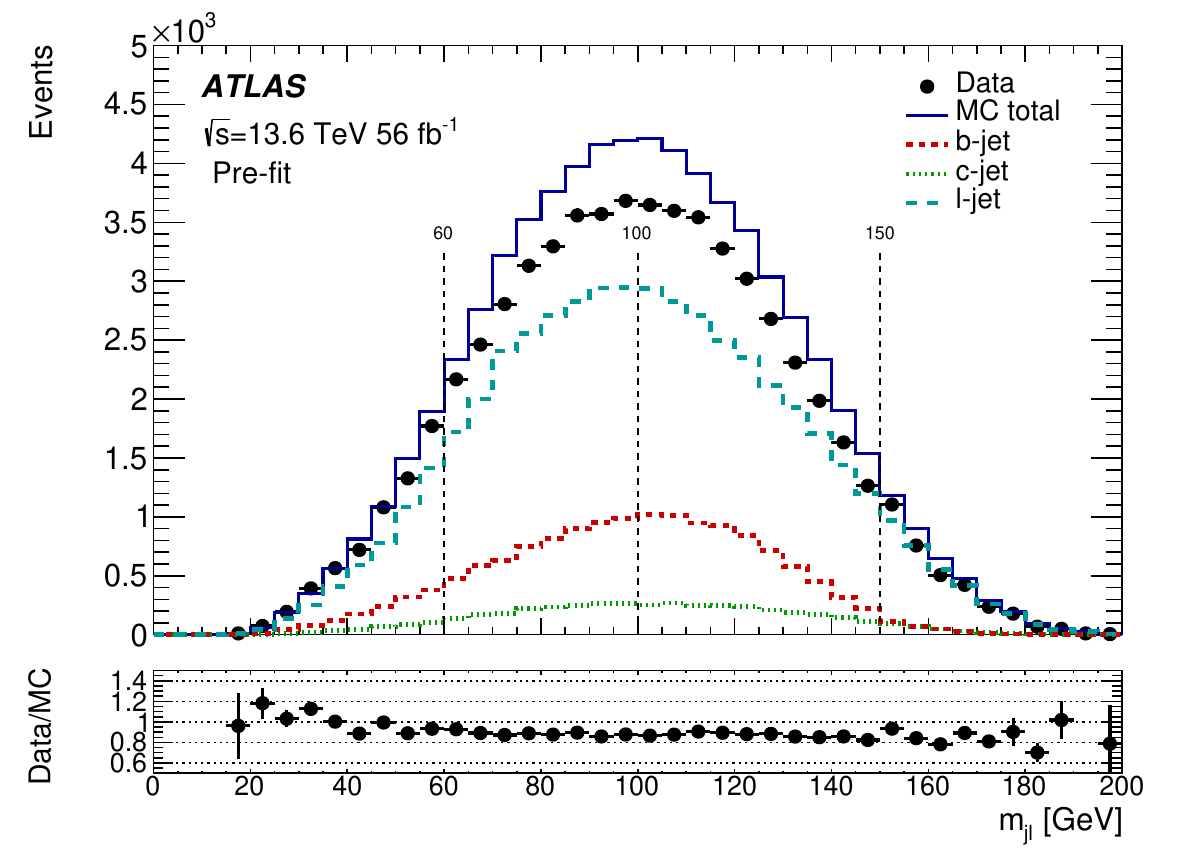} \label{fig:mblprefit}}
\subfigure[Post-fit]{\includegraphics[width=0.45\textwidth]{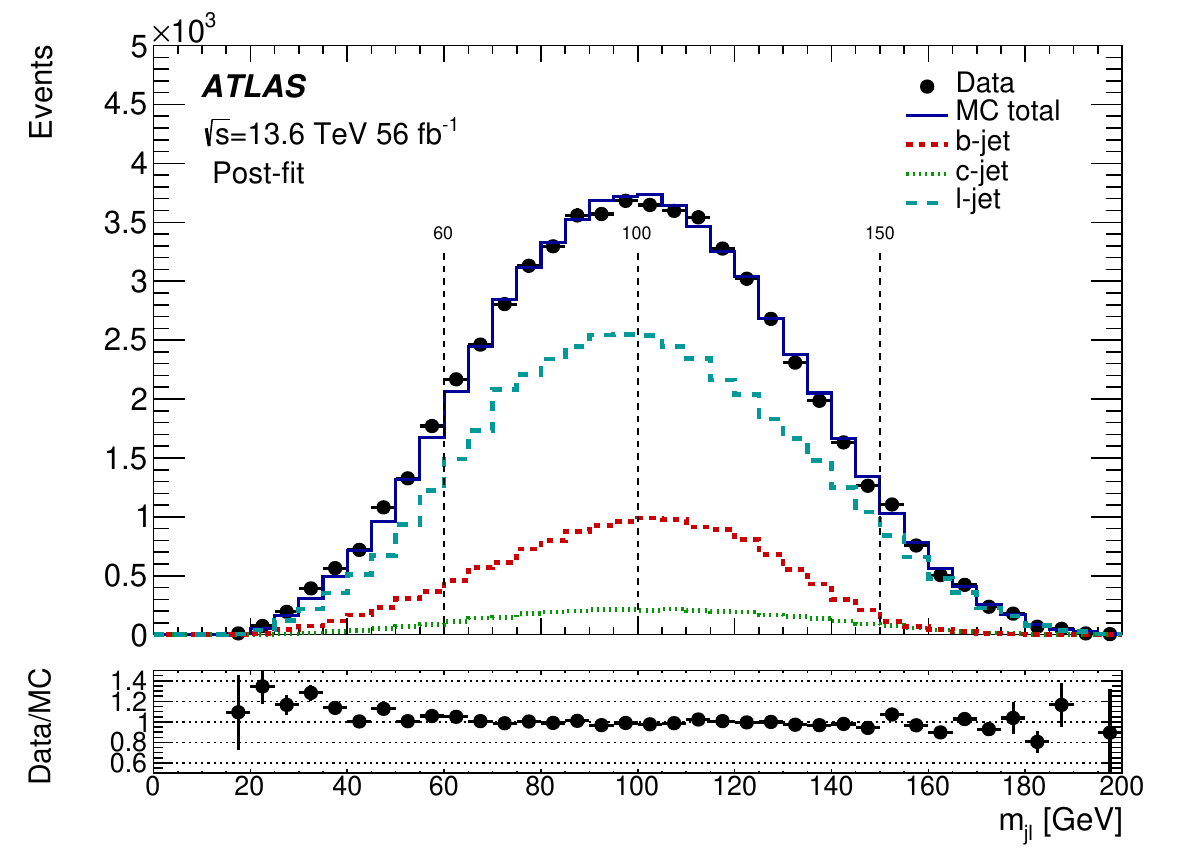} \label{fig:mblpostfit} }
\caption{Distributions of the invariant mass of the charged lepton and leptonic top-jet for events where the hadronic top-jet is tagged in the 0--65\%
interval and the leptonic top-jet and both $W$-jets are untagged.  The  simulation
is split according to the flavour of the jet.  The normalisations and \bjet\ tagging and \cjet\ mistagging
efficiency scale factors, determined in the fit described in this section, are not applied to the simulation in
(a) but are applied in (b). The distributions are shown
without systematic errors.  The dashed vertical lines demarcate the intervals into which the events are separated.}
\label{fig:mbl}
\end{figure}

The $b^{t}(i)$ SFs are mainly constrained by comparing the fraction of
events where the  leptonic top-jet lies within each of the GN2
discriminant intervals with the expectation from MC simulation. Both $W$-jets
are required to be untagged to reduce the rate of
top-quark decay-product misidentification. However, even after this selection
there remains a sizeable fraction of light-flavour jets whose normalisation is
not described well by the simulation, as can be seen in
Figure~\ref{fig:GNprefit}.
In order to normalise the light-flavour jet
contribution, the untagged bin is divided into four intervals according
to the invariant mass of the charged lepton and leptonic top-jet, $m_{jl}$,
with boundaries at 60, 100 and 150~\GeV,
as shown in Figure~\ref{fig:mblprefit}. The quantity $m_{jl}$ is chosen
because it offers discrimination between light-flavour jets and \bjets\ and is
uncorrelated with the GN2 discriminant. The intervals in $m_{jl}$
are labelled with an index $s$, such that $s=-3,-2,-1,0$
for $t=0$ and $s=t$ for $t>0$.  The
number of data events in each bin is denoted by
$N^{s}_{\mathrm{data}}(i)$.

The number of simulated `signal' events in each bin, $N^{s}_{B}(i)$,
corresponds to those events where the leptonic top-jet is a \bjet.
There are two sources of `background' events, defined as events where
the leptonic top-jet is either a light-flavour jet or \cjet, contributing $N^{s}_{L}(i)$
or $N^{s}_{C}(i)$ events respectively. The relative normalisation $L(i)$ of the light-flavour jet
contribution is allowed to vary in the fit in each jet-\pt\ bin. The
relative contribution from \cjets\ is small and set to the expectation from simulation multiplied by $c^{t}(i)$.
The overall normalisation of events, $k_b(i)$, in each jet-\pt\ bin is also allowed
to vary in the fit.

Some events are common to the $\chi_b^2$
and $\chi_c^2$ determinations, specifically those where both $W$-jets are untagged and the
leptonic top-jet lies within the 0--65\% tagging interval. The
statistical correlation between the \bjet\ and \cjet\ SFs is therefore not
precisely determined in the method presented above.  However, the fit uses different properties of the events in common,
so the statistical correlation is expected to be much smaller than the systematic correlation.

The $\chi_b^2$ contribution is given by

\begin{equation*}
\chi_b^2 = \sum_{i=1}^4 \sum_{s=-3}^5   \left[N^{s}_{\mathrm{data}}(i)- k_b(i) \Bigg (L(i)l^{t}(i)N^{s}_L(i) +  c^{t}(i)N^{s}_{C}(i)+  b^{t}(i)N^{s}_{B}(i)\Bigg)\right]^2/N^{s}_{\mathrm{data}}(i).
\end{equation*}

\begin{figure}%
\centering
\subfigure[Pre-fit]{\includegraphics[width=0.45\textwidth]{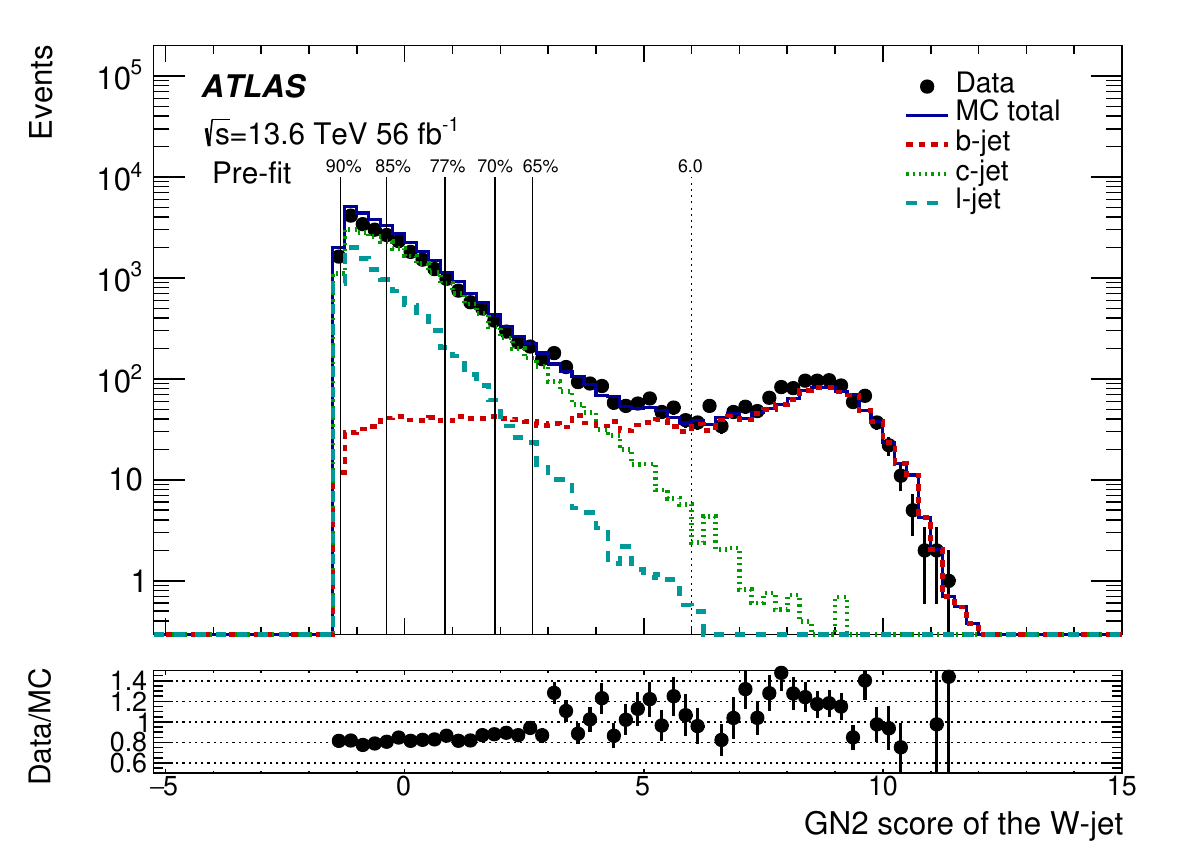}
\label{fig:GNWprefit}}
\subfigure[Post-fit]{\includegraphics[width=0.45\textwidth]{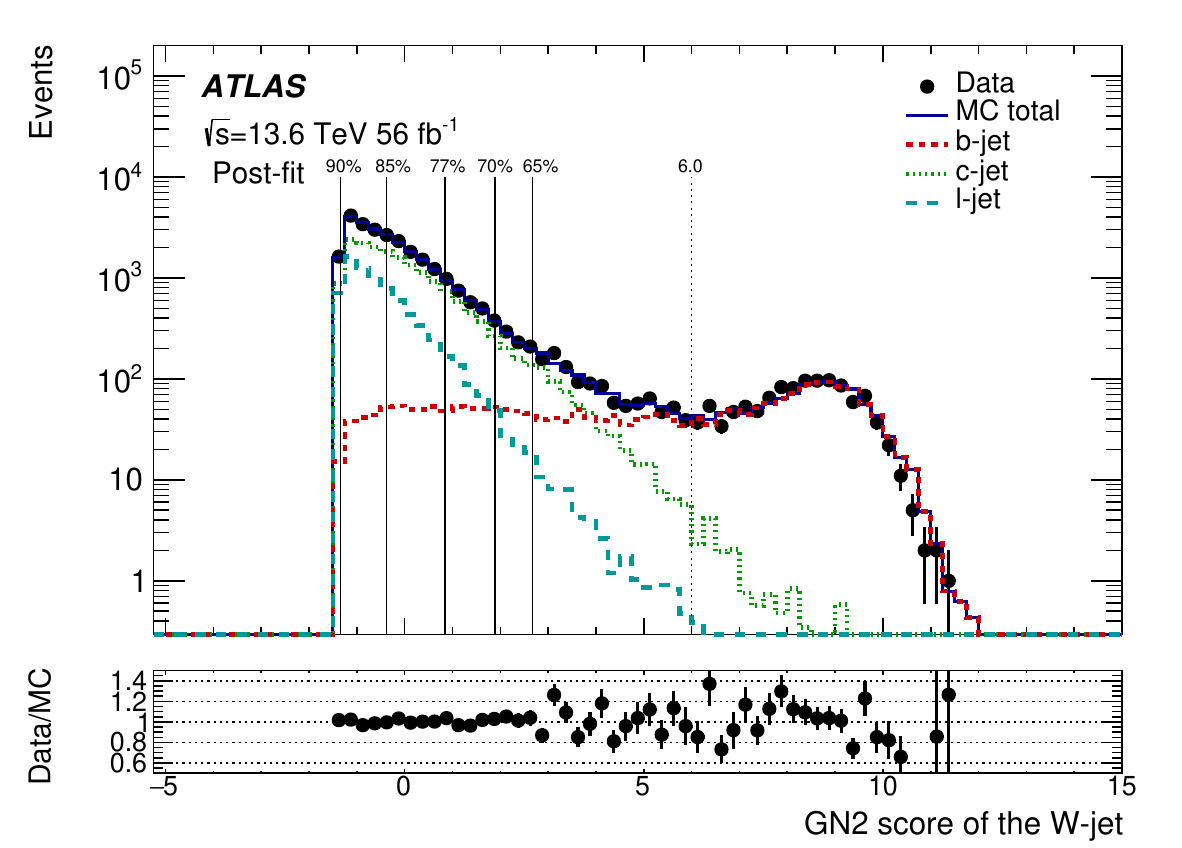}  \label{fig:GNWpostfit} }
\caption{Distributions of the GN2 discriminant of the tagged $W$-jet
for events where the other $W$-jet is
untagged and both top-jets are tagged with a value in the
$0{-}65\%$ interval.  The simulation is split according to the
flavour of the jet. The normalisations and \bjet\ tagging and \cjet\ mistagging
efficiency scale factors, determined in the fit described in this section, are not
applied to the simulation in (a) but are applied in (b). The distributions
are shown without systematic errors. The solid vertical lines demarcate
the tagging intervals in which the measurements of the (mis)tagging
efficiency scale factors are made. The dashed vertical line denotes
the extra splitting in the tightest tagging interval, which is used to
help determine the background normalisation.}
\label{fig:GNW}
\end{figure}

The $c^{t}(i)$ SFs are mainly constrained by comparing the fraction of
events where one of the $W$-jets lies within each of the GN2
discriminant intervals with the expectation from MC simulation. To
simplify the fit, events where both $W$-jets are tagged, making up
0.7\% of the total number of events, are discarded.
In order to reduce the number of wrongly assigned \bjets, both
top-jets must be tagged with the tightest (0--65\%) requirement.   The GN2
discriminant of the tagged $W$-jet is shown in Figure~\ref{fig:GNWprefit}.
It can be seen that the expectation from simulation is dominated by
\cjets, with sizeable contributions from light-flavour jets at low GN2 values and
\bjets\ at high GN2 values.

\begin{figure}%
\centering
{\includegraphics[width=0.45\textwidth]{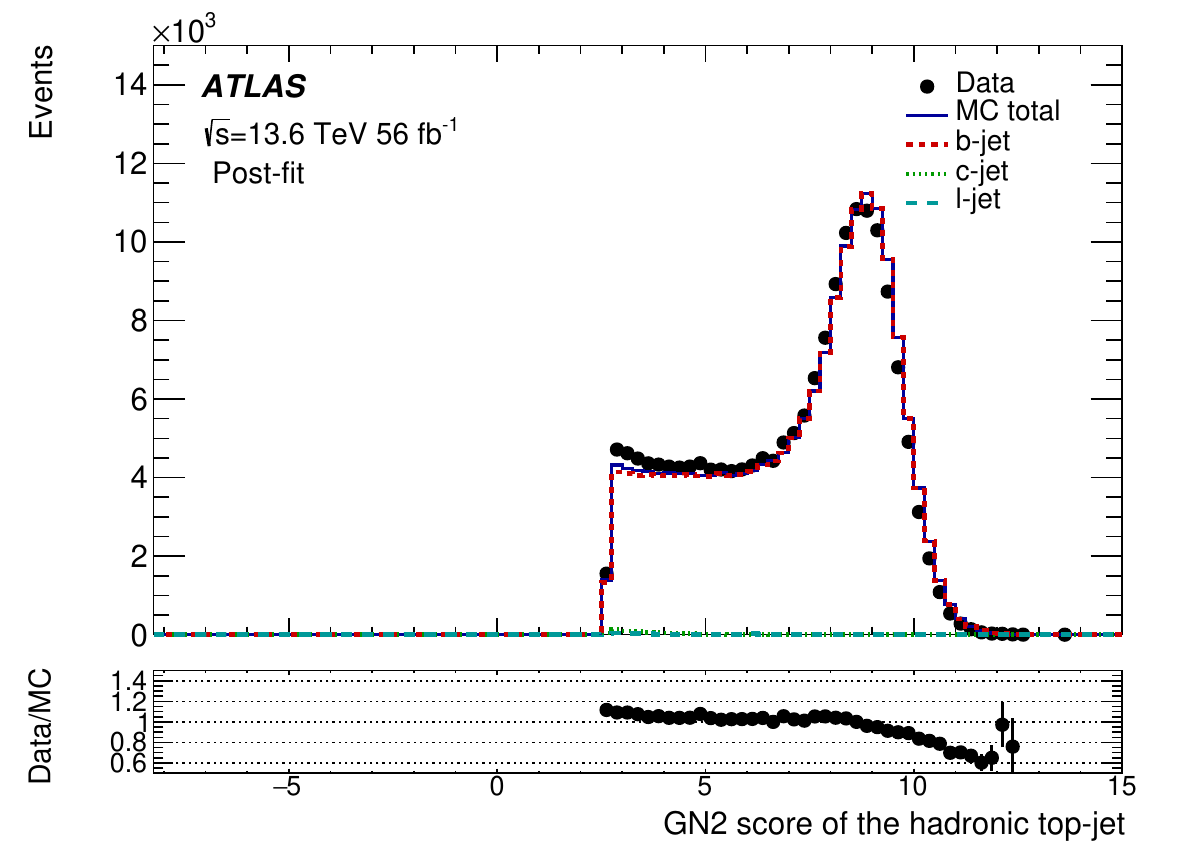}  }
\caption{Distribution of the GN2 discriminant for the 0--65\% tagging interval for
the hadronic top-jet for events where both $W$-jets are untagged.  The  simulation
is split according to the flavour of the jet. The normalisations and \bjet\ tagging and \cjet\ mistagging
efficiency scale factors,  determined in the fit described in this section, are applied to the simulation. The distributions are shown
without systematic errors.}
\label{fig:GN65}
\end{figure}

Events are classified according to the $\pt$ bin of the
tagged $W$-jet (labelled $i$) and the $\pt$ bin of the untagged
$W$-jet (labelled $j$).  In events where both $W$-jets are untagged,
the higher-$\pt$ jet is denoted by $j$, so $j\ge i$.
Contributions from \bjets\ can
arise from misidentified top-jets and also $t\bar{t}$ production with
additional \bjets.  These contributions may not be well described by
simulation, so the relative normalisation is allowed to vary with a
normalisation factor $B$. To determine  $B$, the bin
with the 0--65\% tagging interval ($t=5$) is split into two
for GN2 discriminant values above or below 6.  An index $u$ is used to
label events such that $u=t$ for $t<5$ and $u=5,6$ for $t=5$. The
relative fraction of \bjets\ in the two regions of the 0--65\% tagging interval  is adjusted in simulation to
that observed in data by examining the fraction for the hadronic
top-jet. As can be seen in Figure~\ref{fig:GN65}, this region is
dominated by \bjets. The fraction is evaluated in each of the four
jet-\pt\ intervals and the mean value is taken as the central value.
The uncertainty is taken as the largest deviation from the mean in any of the jet-\pt\ intervals.
The resulting correction either decreases the event weight in the simulation by $(2.4 \pm 1.9)\%$ or
increases it by $(5.1 \pm 1.9)\%$ depending on whether the GN2 discriminant is above or below $6$.
This correction is only applied to the tagged $W$-jet and not to the top-jets.

The number of data events in each bin is $N^{u}_{\mathrm{data}}(i,j)$
and the expected numbers of light-flavour jet, $c$-jet and $b$-jet events are
$N^{u}_{L}(i,j)$, $N^{u}_{C}(i,j)$ and $N^{u}_{B}(i,j)$ respectively.
The overall normalisation of events, $k_W(i,j)$, in each  bin is also
allowed to vary in the fit.

The $c^{t}(i)$ SFs are determined by
minimising the partial $\chi^2$ defined as
\begin{eqnarray*}
\chi_c^2 =\sum_{i=1}^4  \sum_{j=1}^4  \sum_{u=0}^6 \left[N^{u}_{\mathrm{data}}(i,j)- k_W(i,j) l^0(j) \Bigg ( l^t(i)N^{u}_L(i,j)+ c^t(i)N^{u}_{C}(i,j)+ B b^t(i)N^{u}_{B}(i,j) \Bigg)\right]^2/N^{u}_{\mathrm{data}}(i,j).
\end{eqnarray*}
It should be noted that the SF $l^0(j)$ is only applied if the untagged $W$-jet is a light-flavour jet; to simplify the fit, this SF is ignored if the untagged jet is
a \bjet\ or \cjet.

The sum of the efficiencies for the untagged bin and all tagged bins is constrained to unity by including in the total $\chi^2$
the following terms, which have an arbitrary tolerance of $1\%$:

\begin{equation*}
\chi_\epsilon^2 =   \sum_{i=1}^4  \left( 1 -\sum_{t=0}^5 b^{t}(i) \epsilon_{b \mathrm{MC}}^t(i)\right) ^2/(1\%)^2+ \sum_{i=1}^4 \left( 1 -\sum_{t=0}^5 c^{t}(i) \epsilon_{c \mathrm{MC}}^t(i)\right) ^2/(1\%)^2,
\label{eqn:extraterms}
\end{equation*}

where $\epsilon_{b \mathrm{MC}}^t(i)$ and $\epsilon_{c
\mathrm{MC}}^t(i)$ are the MC efficiencies in each tag and $\pt$ bin
for \bjets\ and \cjets\ respectively. With this constraint the
maximum deviation of the sum of the \bjet\ (\cjet) efficiencies from unity
is $0.12\%$ ($0.017\%$) for any \pt\ interval, which is much smaller than the total uncertainty on any of the measured efficiencies.

The SFs $c^{t}(i)$ and $b^{t}(i)$ and the normalisation parameters
$k_b(i)$, $k_W(i,j)$, $L(i)$, $B$ are free parameters in the fit.

The \bjet\ efficiency SFs were also measured in $t\bar{t}$ dileptonic events as reported in Ref.~\cite{btagging}. A relatively simple way of combining these SFs
with those measured in the present analysis is to use the  $t\bar{t}$ dileptonic
measurements, $b^t_{\mathrm{DL}}(i)$ with statistical uncertainty $\Delta b^t_{\mathrm{DL}}(i)$, as a prior.
This is done by adding

\begin{equation*}
\chi_{\mathrm{DL}}^2 = \sum_{i=1}^4 \sum_{t=0}^5    [\, (b^t_{\mathrm{DL}}(i) - b^t(i)) /\Delta b^t_{\mathrm{DL}}(i) \, ]^2
\label{eqn:extraterms2}
\end{equation*}

to the $\chi^2$ defined in Eq.~(\ref{eqn:chi2}).


%
\subsection{Systematic uncertainties}
\label{sec:systematics}
Several sources of experimental and MC modelling systematic
uncertainty affecting the efficiency SFs are considered. The
uncertainty from each source is evaluated by and repeating the fit after
making the corresponding change to the MC model to produce new templates
for the semileptonic or  shifting the dileptonic data points for the combined fit.  Each systematic error
source is thus correlated between the tagging and jet-$\pt$ intervals,
between bottom tagging and charm mistagging SFs and between the
semileptonic and dileptonic analyses. The total systematic
uncertainty for each interval is evaluated by adding the individual
uncertainties in quadrature. The uncertainties of the \bjet\ SFs are
largest in the $90{-}100\%$ interval, where the light-flavour jet background
is the largest, while the \cjet\ SF uncertainties are largest in the $0{-}65\%$
interval, where the \bjet\ background is the largest.

The uncertainties in the jet energy scale and resolution
are derived by combining information from data and
simulation~\cite{JETM-2018-05}, resulting in many independent model variations.
In this analysis, 11 such variations are found to have a significant
effect: the largest leads to changes of up to
15\% (8\%) in the \bjet\ (\cjet) SFs.
The uncertainty due to pile-up modelling~\cite{ATL-PHYS-PUB-2019-026}
leads to changes of up to 5\% (11\%) in the  \bjet\ (\cjet) SFs.

The uncertainties in the light-flavour jet mistagging SFs are accounted
for by changing the SFs by each component of the total uncertainty,
and where appropriate including the correlation between this and the uncertainty used in
the \cjet\ SF determination. The most important uncorrelated
uncertainties originate from
changing the model used for
the simulation of $Z$+jets, giving uncertainties of up to 1\% (5\%), and the MC sample sizes, giving
uncertainties of up to 2\% (5\%) in the \bjet\ (\cjet) SFs~\cite{FTAG-2019-02}.
The uncertainty
in the \bjet\ fraction reweighting in the tightest tagging interval,
described in Section~\ref{sec:method}, leads to a maximum uncertainty
of 2\% in the \cjet\ SF.

Uncertainties from the modelling of  $t\bar{t}$ are
estimated by replacing the nominal MC sample by alternative MC
samples.  The uncertainty due to the choice of parton shower and
hadronisation model is estimated by replacing the standard
\PYTHIA[8] model with \HERWIG.
This is known to have a large effect in the differential
distributions~\cite{ATL-PHYS-PUB-2025-047} and it is the largest source
of uncertainty in this analysis, leading to changes of up to
26\% (13\%) in the \bjet\ (\cjet) SFs.

The uncertainty from the modelling of ISR/FSR is assessed by increasing the
radiation by doubling the \hdamp\
parameter~\cite{ATL-PHYS-PUB-2016-020}. The resulting
uncertainty in the \bjet\ (\cjet) SFs is at most 18\% (21\%).
Finally, the effect of the matrix-element matching
uncertainty on the \bjet\ (\cjet) SFs is at most 14\% (13\%).

Uncertainties in the normalisation of backgrounds other than
$t\bar{t}$ are accounted for by scaling the MC expectation by $\pm
25$\%, which is approximately the maximum difference seen
between data and simulation in dedicated measurements~\cite{TOPQ-2012-20,TOPQ-2018-07}.
The resulting uncertainty in the \bjet\  (\cjet)  SFs is at most 13\% (1\%).

%
%
%


%
\subsection{Results}
\label{sec:results}

\begin{figure}[htbp]
\centering
\includegraphics[scale = 0.4]{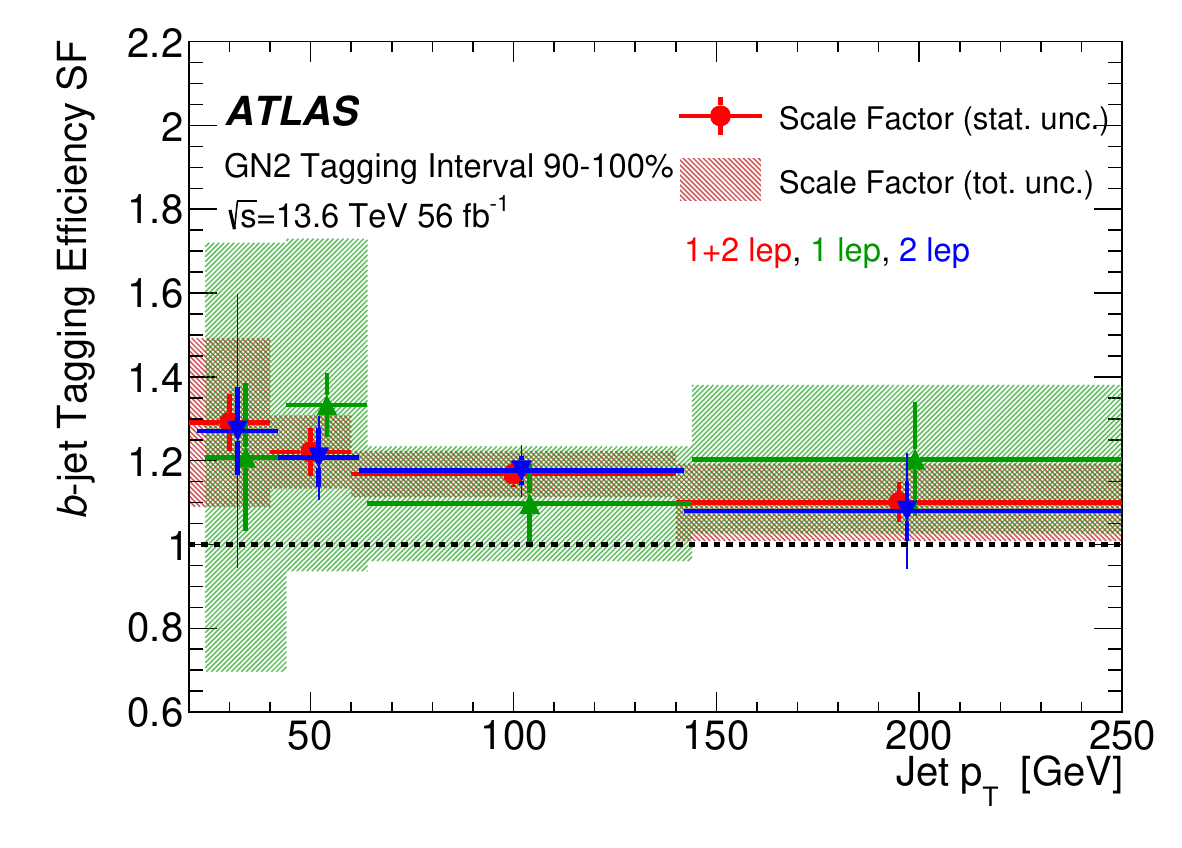}
\includegraphics[scale = 0.4]{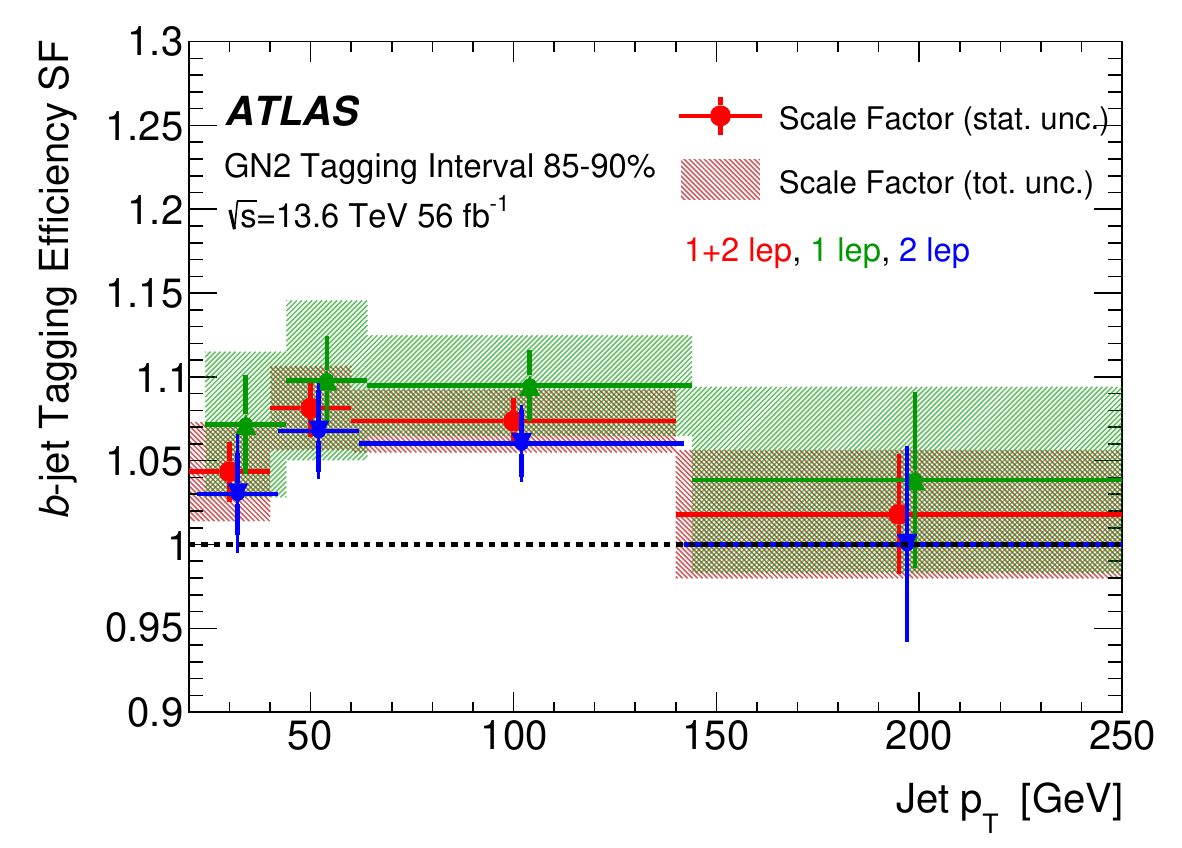} \\
\includegraphics[scale = 0.4]{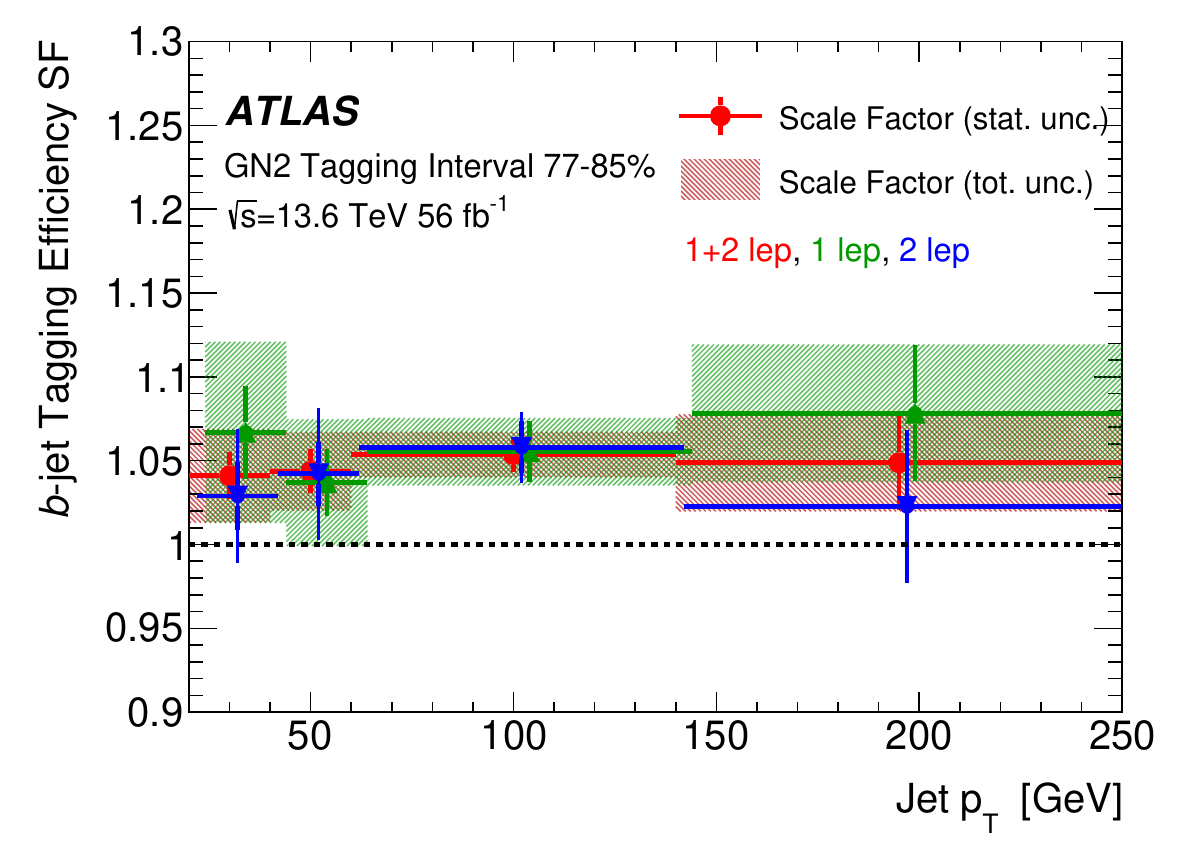}
\includegraphics[scale = 0.4]{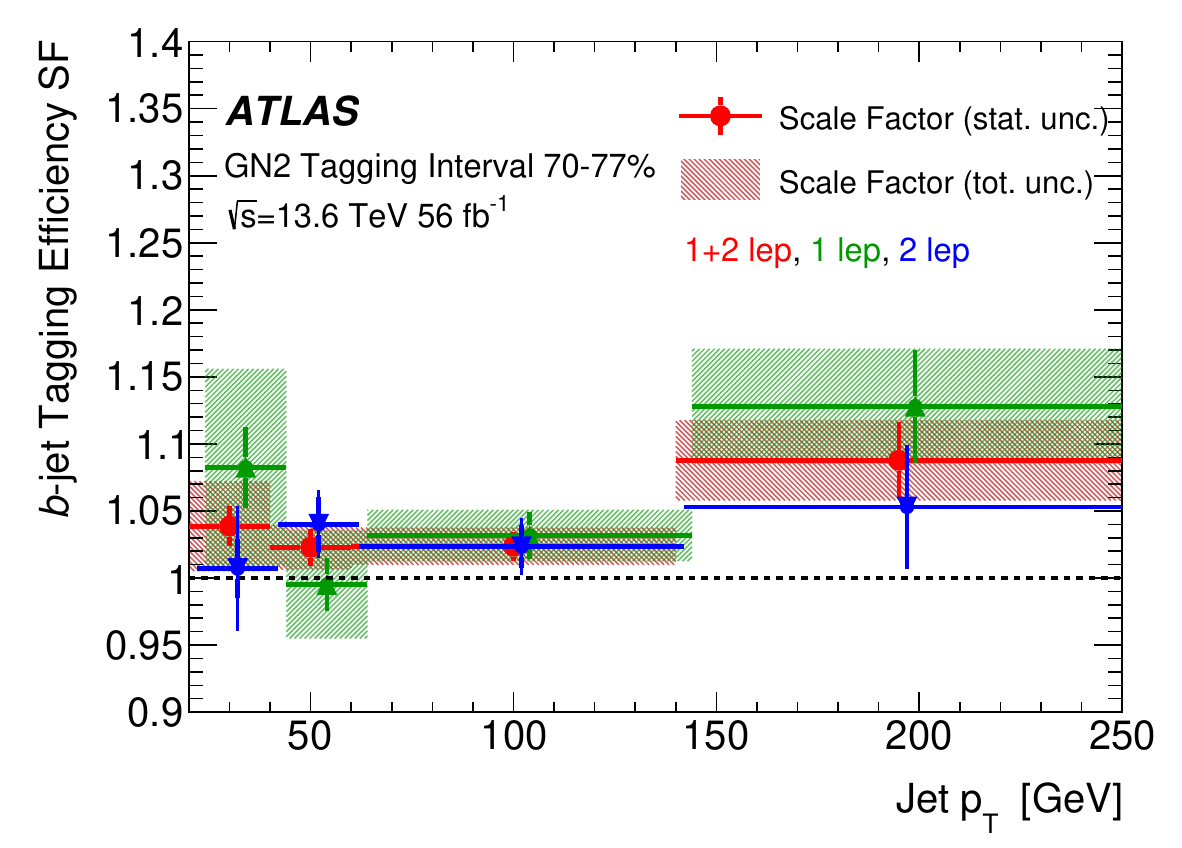} \\
\includegraphics[scale = 0.4]{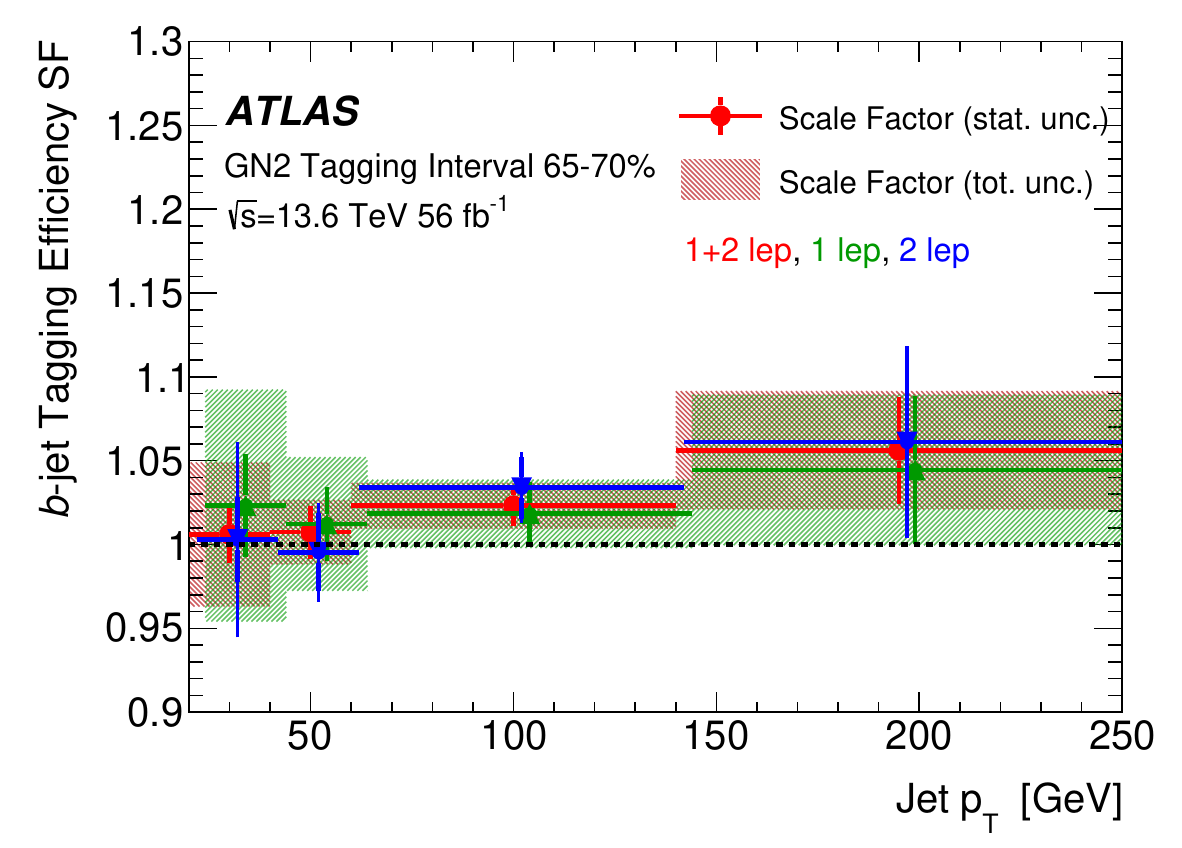}
\includegraphics[scale = 0.4]{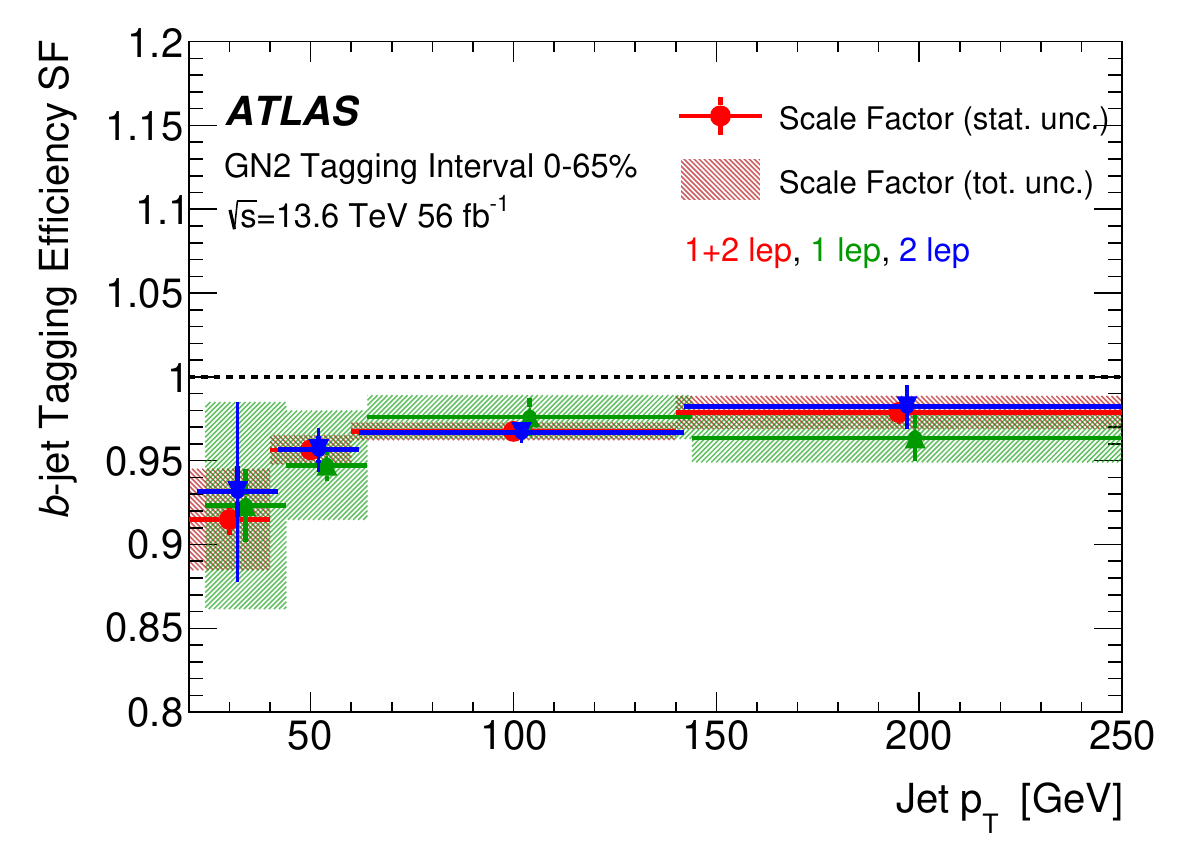}
\caption{ The \bjet\ pseudo-continuous tagging efficiency SFs measured in  semileptonic (1~lep) $t\bar{t}$
events   shown as a function of jet \pt\ for  the six
tagging intervals. Also shown are the measurements made in dileptonic (2~lep) $t\bar{t}$ events and the combination (1+2~lep).
The SFs are shown for  the $\PYTHIA[8]$\,+\,\EVTGEN fragmentation model.}
\label{fig:BSFs}
\end{figure}

The measured  \bjet\ pseudo-continuous tagging efficiency
SFs together with the total, statistical and systematic uncertainties
are listed in Table~\ref{tab:BSFs} and shown in Figure~\ref{fig:BSFs}
for the six tagging intervals. The fitted normalisation factors $k_b(i)$
lie in the range 0.75 to 0.83, and the fitted light-flavour jet background normalisation
factors $L(i)$ lie between 0.97 and 1.16.   The contributions to the
systematic uncertainty are listed in Table~\ref{tab:Berrors}.  It can
be seen that the largest uncertainties arise from $t\bar{t}$ modelling
and jet reconstruction and are largest in the untagged interval, where
the light-flavour jet background is large.  The SFs do not have a strong jet-\pt\
dependence.  The total uncertainty varies from 2\% to 44\% and is
dominated by the systematic uncertainty except in the highest \pt\
bin.

Also shown in Figure~\ref{fig:BSFs} are the measurements of
the \ttbar\ dileptonic \bjet\ tagging efficiency SFs~\cite{btagging}. It can be seen
that the two measurements are in good agreement. The combination of
the two measurements, obtained using the method described in Section~\ref{sec:method}, is also shown.  As is also seen in the individual
measurements, there is no clear \pt\ dependence, but the SF differs
from unity in some tagging intervals.

\begin{figure}[htbp]
\centering
\includegraphics[scale = 0.4]{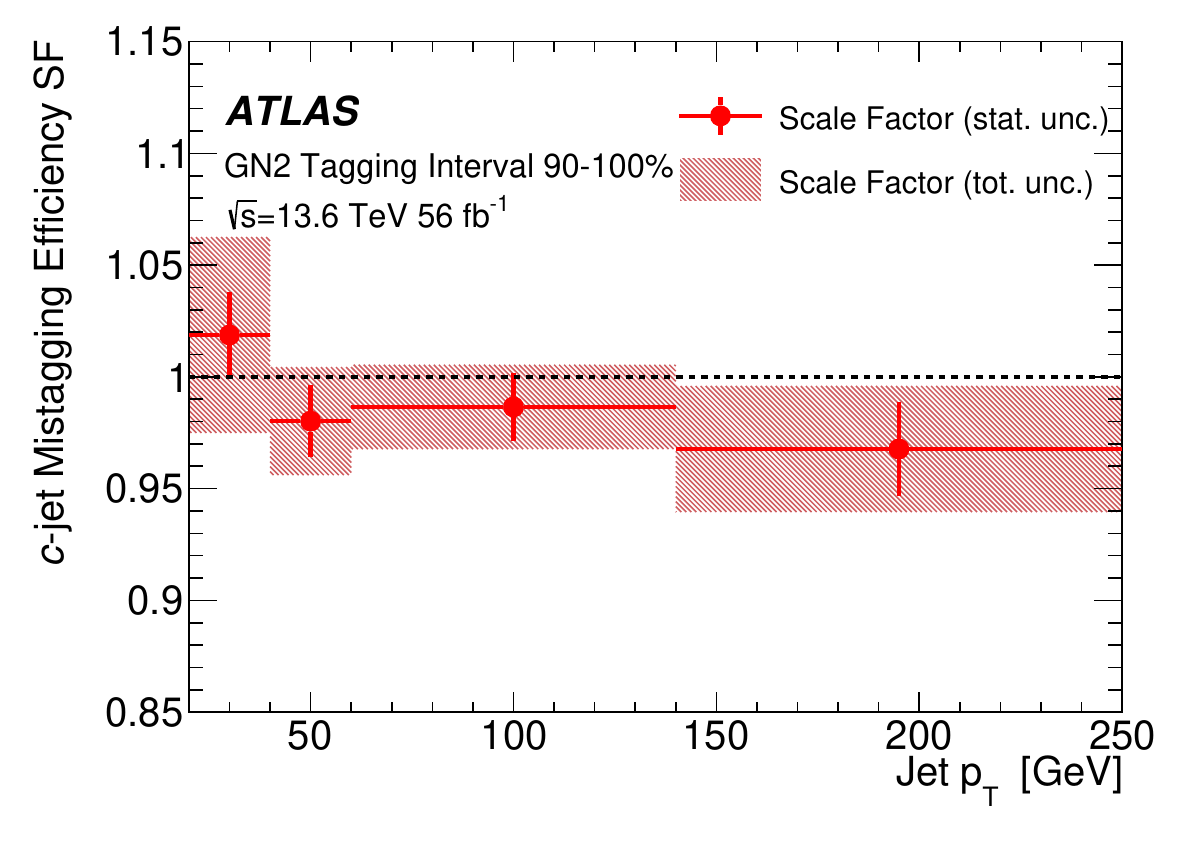}
\includegraphics[scale = 0.4]{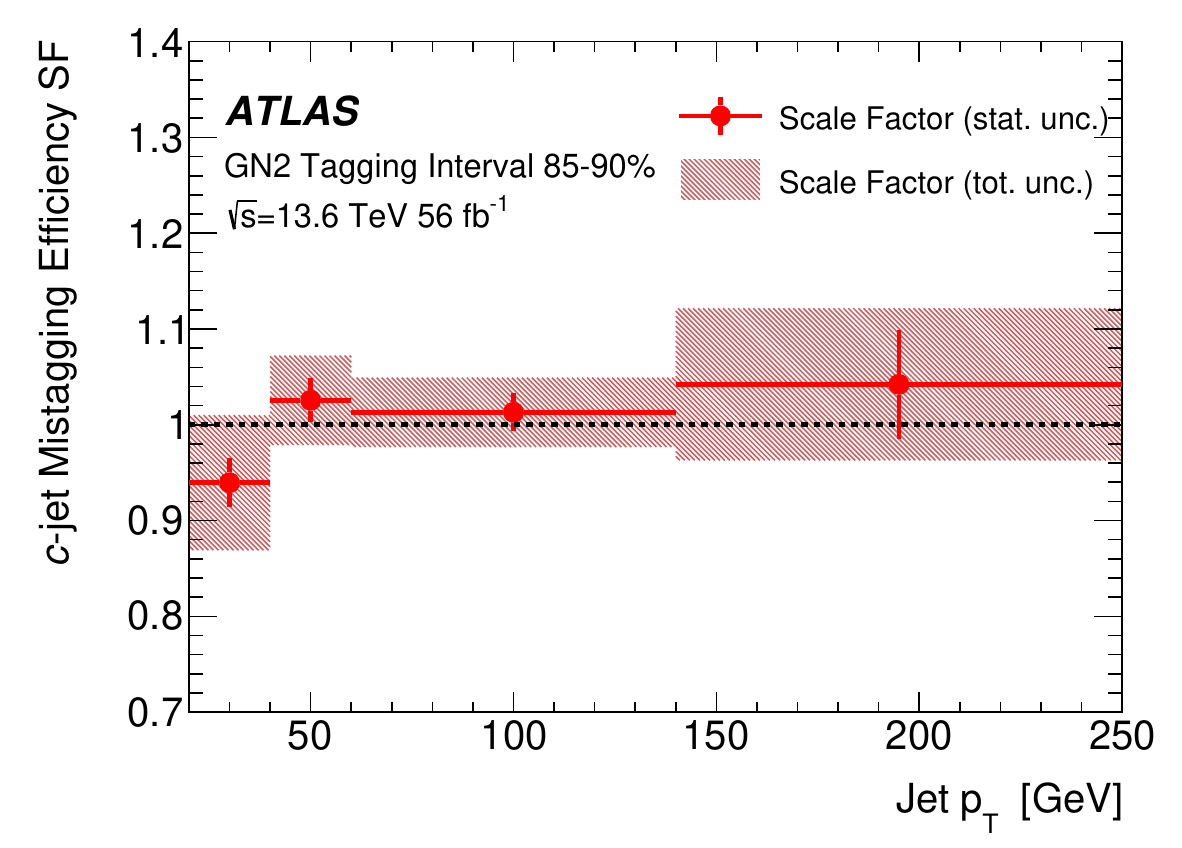} \\
\includegraphics[scale = 0.4]{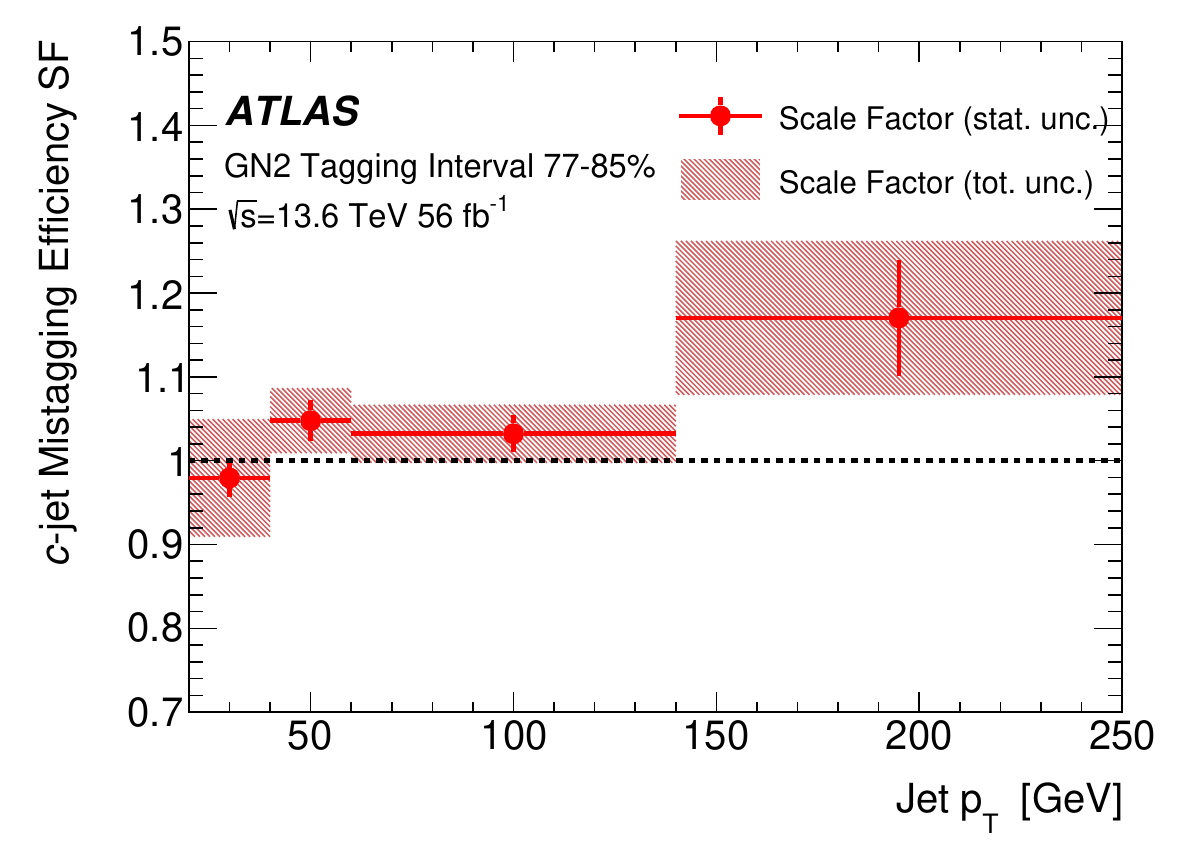}
\includegraphics[scale = 0.4]{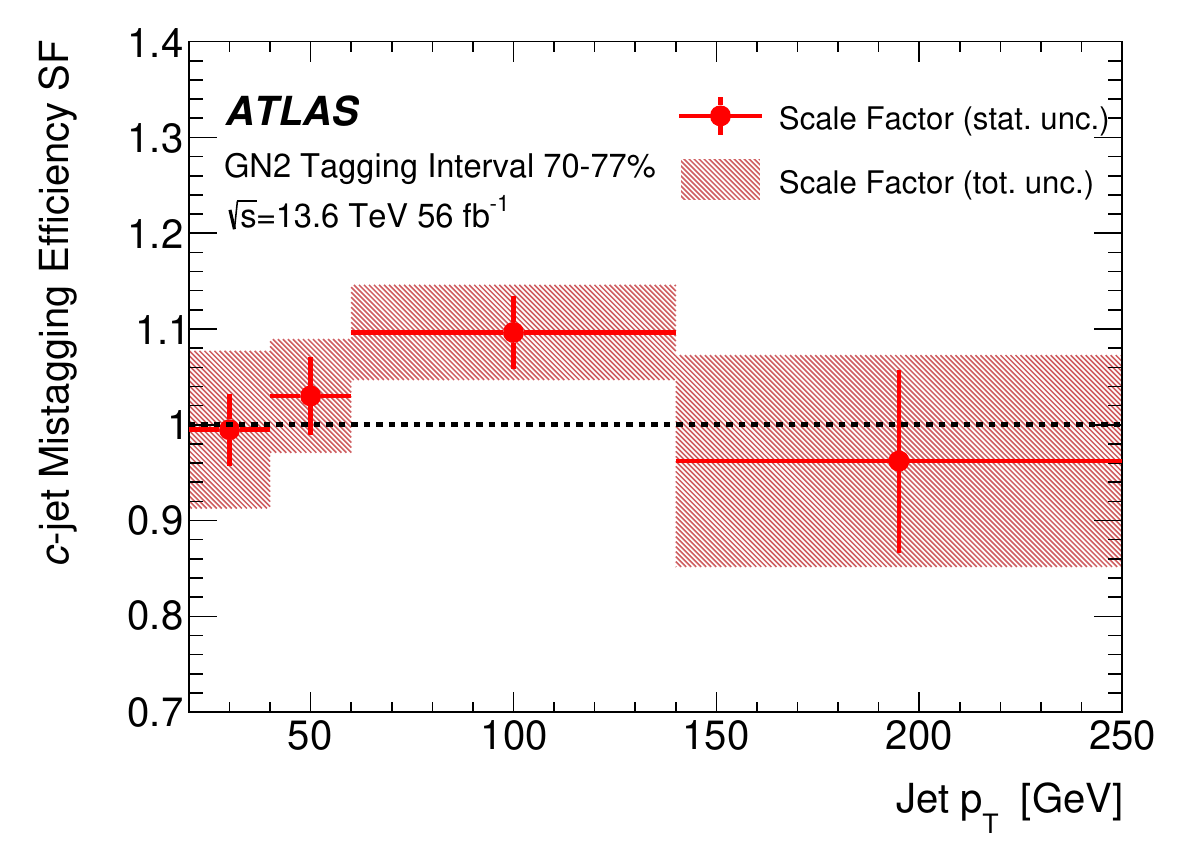} \\
\includegraphics[scale = 0.4]{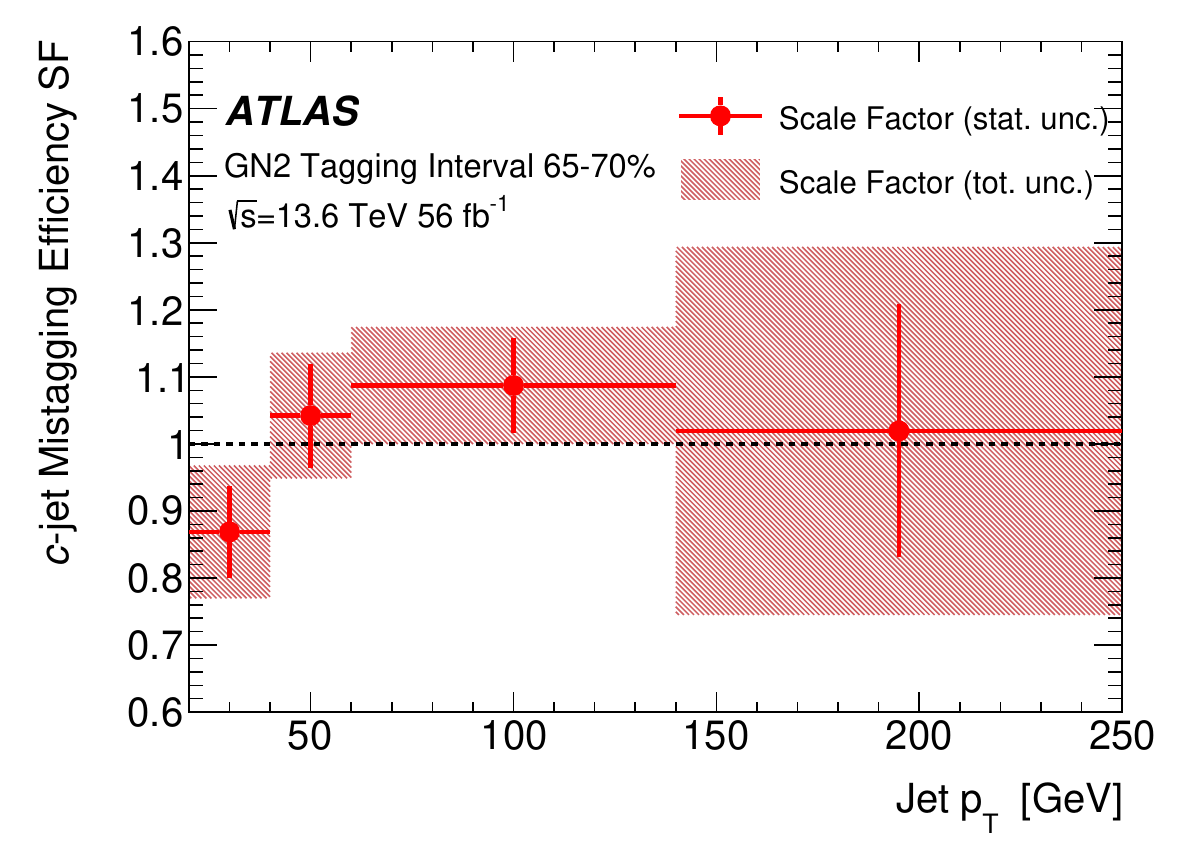}
\includegraphics[scale = 0.4]{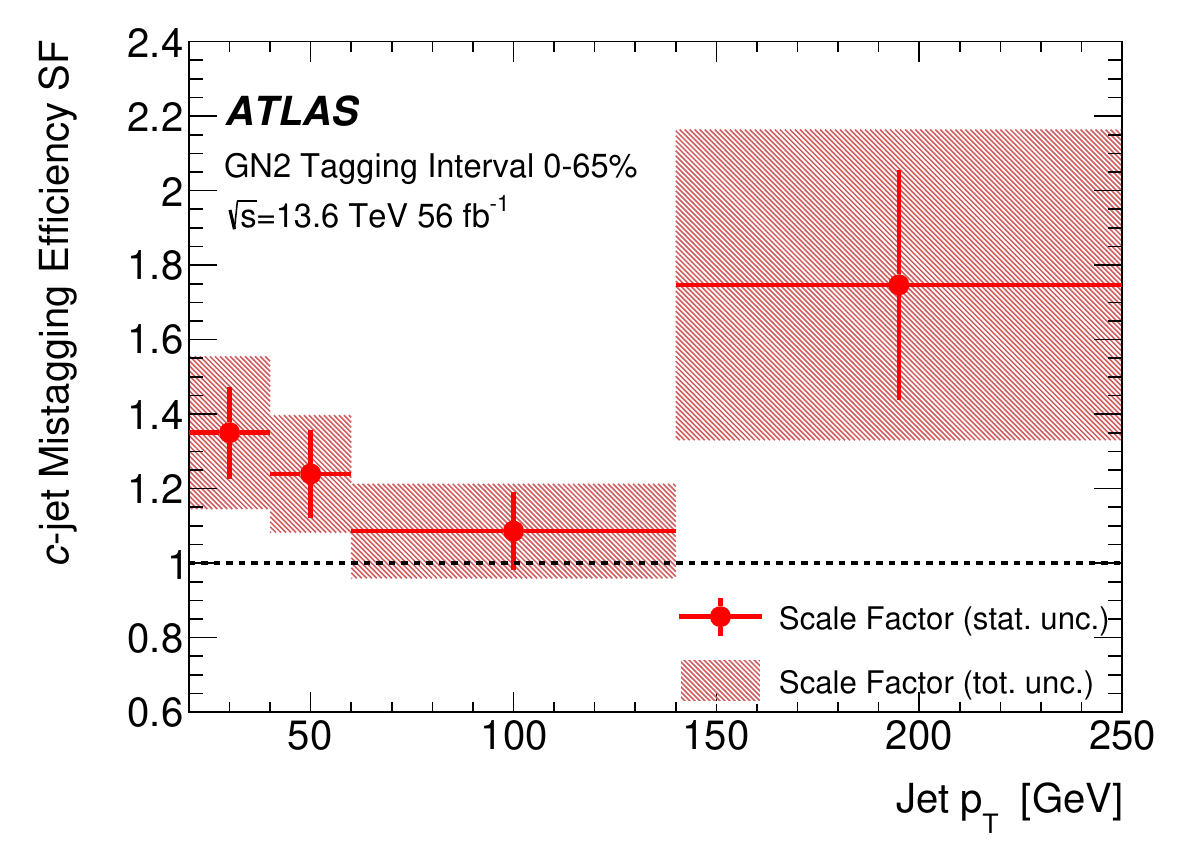}
\caption{ The \cjet\ pseudo-continuous mistagging efficiency SFs measured in  semileptonic $t\bar{t}$ events
shown as a function of jet \pt\ for the six
tagging intervals.  The SFs are shown for  the $\PYTHIA[8]$\,+\,\EVTGEN fragmentation model.}
\label{fig:CSFs}
\end{figure}

The measured \cjet\ pseudo-continuous mistagging efficiency SFs
together with the total, statistical and systematic uncertainties are
listed in Table~\ref{tab:CSFs} and shown in Figure~\ref{fig:CSFs} for
the six tagging intervals.
The fitted normalisation factors $k_W(i,j)$ lie in the range 0.75 to 0.84 and the \bjet\
background normalisation factor is $B=1.56 \pm 0.05\,\text{(stat.)} \pm 0.21\,\text{(syst.)}$.
The contributions to the systematic uncertainty are listed in Table~\ref{tab:Cerrors}.
The SFs do not have a strong jet-\pt\ dependence.
In the tightest tagging interval the efficiency is higher in data than in MC simulation.
The total uncertainty varies from 2\% to 41\%.

\begin{table}[htbp]
\centering
\caption{The measured \bjet\ pseudo-continuous tagging efficiency SFs measured in  semileptonic $t\bar{t}$ events, listed with the total, relative statistical and relative systematic uncertainties for each tagging and jet-$\pt$ interval. The SFs are listed for  the $\PYTHIA[8]$\,+\,\EVTGEN fragmentation model.}
\vspace{0.2cm}
\begin{tabular}{cc|c|ccc}
&   &   & \multicolumn{3}{c}{SF uncertainty} \\
Tagging interval  & Jet $\pt$ range [$\GeV$] & SF & Total & Stat. & Syst. \\ \hline
%

90--100\% & 20--40&1.206 & 43\% & 14\% & 41\% \\
90--100\% & 40--60&1.335 & 31\% & 13\% & 28\% \\
90--100\% & 60--140&1.096 & 15\% & 8.5\% & 12\% \\
90--100\% & 140--250&1.200 & 14\% & 11\% & 8.3\% \\
85--90\% & 20--40&1.068 & 4.8\% & 3.0\% & 3.8\% \\
85--90\% & 40--60&1.092 & 5.1\% & 2.8\% & 4.3\% \\
85--90\% & 60--140&1.095 & 2.4\% & 2.0\% & 1.2\% \\
85--90\% & 140--250&1.035 & 4.8\% & 4.6\% & 1.3\% \\
77--85\% & 20--40&1.063 & 6.4\% & 3.0\% & 5.7\% \\
77--85\% & 40--60&1.031 & 5.1\% & 2.6\% & 4.4\% \\
77--85\% & 60--140&1.055 & 2.5\% & 1.8\% & 1.7\% \\
77--85\% & 140--250&1.075 & 4.0\% & 3.8\% & 1.2\% \\
70--77\% & 20--40&1.079 & 8.1\% & 3.2\% & 7.5\% \\
70--77\% & 40--60&0.990 & 5.1\% & 2.7\% & 4.4\% \\
70--77\% & 60--140&1.032 & 2.1\% & 1.8\% & 1.1\% \\
70--77\% & 140--250&1.125 & 3.9\% & 3.8\% & 0.9\% \\
65--70\% & 20--40&1.020 & 8.7\% & 3.4\% & 8.0\% \\
65--70\% & 40--60&1.007 & 5.1\% & 2.9\% & 4.2\% \\
65--70\% & 60--140&1.018 & 2.3\% & 2.0\% & 1.2\% \\
65--70\% & 140--250&1.041 & 4.5\% & 4.3\% & 1.1\% \\
0--65\% & 20--40&0.920 & 8.5\% & 2.8\% & 8.0\% \\
0--65\% & 40--60&0.942 & 4.6\% & 2.1\% & 4.0\% \\
0--65\% & 60--140&0.976 & 1.7\% & 1.3\% & 1.1\% \\
0--65\% & 140--250&0.961 & 1.7\% & 1.5\% & 0.8\% \\


\end{tabular}
\label{tab:BSFs}
\end{table}

\begin{table}[htbp]
\centering
\caption{The contributions to the
\bjet\ pseudo-continuous tagging efficiency SF systematic uncertainties measured in  semileptonic $t\bar{t}$ events, listed as a function of jet \pt\ for  six
tagging intervals. Included are the relative uncertainties related to the $t\bar{t}$ modelling, the jets and \MET, the light-flavour jet SF and all other sources. }
\vspace{0.2cm}
\begin{tabular}{cc|cccc}
&   & \multicolumn{4}{c}{SF systematic uncertainty} \\
Tagging interval  & Jet $\pt$ range [$\GeV$] &  $t\bar{t}$ mod. &  Jet \& \MET & Light tag   & Other \\  \hline

90--100\% & 20--40&32\% & 22\% & 2.4\%  & 13\% \\
90--100\% & 40--60&22\% & 17\% & 1.8\%  & 8.1\% \\
90--100\% & 60--140&9.1\% & 4.1\% & 2.7\%  & 5.5\% \\
90--100\% & 140--250&5.6\% & 5.8\% & 1.3\%  & 1.4\% \\
85--90\% & 20--40&2.7\% & 1.6\% & 0.9\%  & 1.9\% \\
85--90\% & 40--60&2.1\% & 3.5\% & 0.5\%  & 1.2\% \\
85--90\% & 60--140&0.9\% & 0.6\% & 0.0\%  & 0.6\% \\
85--90\% & 140--250&1.0\% & 0.7\% & 0.0\%  & 0.5\% \\
77--85\% & 20--40&4.4\% & 2.6\% & 0.2\%  & 2.4\% \\
77--85\% & 40--60&3.6\% & 2.2\% & 0.2\%  & 1.2\% \\
77--85\% & 60--140&1.5\% & 0.5\% & 0.2\%  & 0.6\% \\
77--85\% & 140--250&1.1\% & 0.5\% & 0.0\%  & 0.2\% \\
70--77\% & 20--40&5.5\% & 4.5\% & 0.4\%  & 2.4\% \\
70--77\% & 40--60&3.2\% & 2.7\% & 0.2\%  & 1.2\% \\
70--77\% & 60--140&0.9\% & 0.4\% & 0.2\%  & 0.5\% \\
70--77\% & 140--250&0.6\% & 0.6\% & 0.0\%  & 0.1\% \\
65--70\% & 20--40&6.5\% & 3.9\% & 0.6\%  & 2.4\% \\
65--70\% & 40--60&3.3\% & 2.3\% & 0.2\%  & 1.1\% \\
65--70\% & 60--140&0.9\% & 0.5\% & 0.2\%  & 0.5\% \\
65--70\% & 140--250&1.0\% & 0.5\% & 0.0\%  & 0.1\% \\
0--65\% & 20--40&6.4\% & 4.2\% & 0.6\%  & 2.3\% \\
0--65\% & 40--60&3.1\% & 2.4\% & 0.2\%  & 1.1\% \\
0--65\% & 60--140&0.9\% & 0.5\% & 0.3\%  & 0.5\% \\
0--65\% & 140--250&0.6\% & 0.5\% & 0.0\%  & 0.1\% \\


\end{tabular}
\label{tab:Berrors}
\end{table}

\begin{table}[htbp]
\centering
\caption{The measured \cjet\ pseudo-continuous mistagging efficiency SFs measured in  semileptonic $t\bar{t}$ events, listed with the total, relative statistical and relative systematic uncertainties for each tagging and jet-$\pt$ interval. The SFs are listed for  the $\PYTHIA[8]$\,+\,\EVTGEN fragmentation model.}
\vspace{0.2cm}
\begin{tabular}{cc|c|ccc}
&   &   & \multicolumn{3}{c}{SF uncertainty} \\
Tagging interval  & Jet $\pt$ range [$\GeV$] & SF & Total & Stat. & Syst. \\ \hline

90--100\% & 20--40&1.019 & 3.9\% & 1.9\% & 3.4\% \\
90--100\% & 40--60&0.980 & 2.7\% & 1.7\% & 2.1\% \\
90--100\% & 60--140&0.987 & 1.9\% & 1.6\% & 1.2\% \\
90--100\% & 140--250&0.968 & 3.0\% & 2.2\% & 2.0\% \\
85--90\% & 20--40&0.939 & 7.1\% & 2.7\% & 6.5\% \\
85--90\% & 40--60&1.026 & 5.2\% & 2.3\% & 4.7\% \\
85--90\% & 60--140&1.013 & 3.7\% & 2.0\% & 3.1\% \\
85--90\% & 140--250&1.042 & 9.5\% & 5.7\% & 7.7\% \\
77--85\% & 20--40&0.979 & 6.1\% & 2.4\% & 5.6\% \\
77--85\% & 40--60&1.048 & 4.5\% & 2.4\% & 3.7\% \\
77--85\% & 60--140&1.032 & 4.1\% & 2.1\% & 3.5\% \\
77--85\% & 140--250&1.170 & 7.3\% & 6.1\% & 4.1\% \\
70--77\% & 20--40&0.995 & 9.8\% & 3.8\% & 9.0\% \\
70--77\% & 40--60&1.030 & 6.2\% & 4.1\% & 4.7\% \\
70--77\% & 60--140&1.096 & 4.8\% & 3.5\% & 3.3\% \\
70--77\% & 140--250&0.962 & 12\% & 10\% & 6.4\% \\
65--70\% & 20--40&0.868 & 12\% & 8.1\% & 9.2\% \\
65--70\% & 40--60&1.042 & 9.5\% & 7.6\% & 5.7\% \\
65--70\% & 60--140&1.086 & 15\% & 6.6\% & 14\% \\
65--70\% & 140--250&1.019 & 32\% & 18.9\% & 25\% \\
0--65\% & 20--40&1.342 & 17\% & 9.6\% & 14\% \\
0--65\% & 40--60&1.234 & 17\% & 11\% & 12\% \\
0--65\% & 60--140&1.074 & 12\% & 10\% & 6.3\% \\
0--65\% & 140--250&1.740 & 32\% & 18\% & 26\% \\


%
\end{tabular}
\label{tab:CSFs}
\end{table}

\begin{table}[htbp]
\centering
\caption{The contributions to the
\cjet\ pseudo-continuous mistagging efficiency SF systematic uncertainties measured in  semileptonic $t\bar{t}$ events, listed as a function of jet \pt\ for  six
tagging intervals. Included are the relative uncertainties related to the $t\bar{t}$ modelling, the jets and \MET, the light-flavour jet SF, the \bjet\
SF, and all other sources. }
\vspace{0.2cm}
\begin{tabular}{cc|ccccc}
&   & \multicolumn{5}{c}{SF systematic uncertainty} \\
Tagging interval  & Jet $\pt$ range [$\GeV$] &  $t\bar{t}$ mod. &  Jet \& \MET & Light tag & $b$-tag  & Other \\  \hline

90--100\% & 20--40&1.3\% & 2.5\% & 1.9\% & 0.0\%  & 0.0\% \\
90--100\% & 40--60&1.2\% & 1.3\% & 1.0\% & 0.0\%  & 0.0\% \\
90--100\% & 60--140&0.8\% & 0.5\% & 0.7\% & 0.0\%  & 0.0\% \\
90--100\% & 140--250&1.7\% & 0.5\% & 0.8\% & 0.0\%  & 0.3\% \\
85--90\% & 20--40&1.9\% & 4.8\% & 4.0\% & 0.0\%  & 0.3\% \\
85--90\% & 40--60&2.9\% & 3.0\% & 2.1\% & 0.0\%  & 0.5\% \\
85--90\% & 60--140&1.9\% & 1.7\% & 1.8\% & 0.0\%  & 0.0\% \\
85--90\% & 140--250&6.8\% & 1.9\% & 2.4\% & 0.0\%  & 1.8\% \\
77--85\% & 20--40&2.3\% & 4.2\% & 3.0\% & 0.0\%  & 0.0\% \\
77--85\% & 40--60&2.9\% & 1.7\% & 1.6\% & 0.0\%  & 0.0\% \\
77--85\% & 60--140&2.8\% & 1.6\% & 1.3\% & 0.0\%  & 0.0\% \\
77--85\% & 140--250&3.4\% & 1.3\% & 1.6\% & 0.0\%  & 0.7\% \\
70--77\% & 20--40&7.3\% & 3.3\% & 4.1\% & 0.1\%  & 0.3\% \\
70--77\% & 40--60&2.8\% & 2.8\% & 2.5\% & 0.0\%  & 0.0\% \\
70--77\% & 60--140&2.3\% & 0.9\% & 2.1\% & 0.0\%  & 0.4\% \\
70--77\% & 140--250&5.6\% & 1.6\% & 2.5\% & 0.0\%  & 0.0\% \\
65--70\% & 20--40&6.7\% & 4.3\% & 4.3\% & 0.3\%  & 1.6\% \\
65--70\% & 40--60&3.3\% & 2.8\% & 2.8\% & 0.2\%  & 2.5\% \\
65--70\% & 60--140&13\% & 1.5\% & 3.6\% & 0.2\%  & 1.7\% \\
65--70\% & 140--250&18\% & 12\% & 8.1\% & 0.2\%  & 11\% \\
0--65\% & 20--40&9.3\% & 9.3\% & 2.5\% & 1.8\%  & 3.3\% \\
0--65\% & 40--60&9.4\% & 7.2\% & 1.8\% & 1.9\%  & 0.9\% \\
0--65\% & 60--140&2.6\% & 3.2\% & 4.3\% & 2.0\%  & 0.7\% \\
0--65\% & 140--250&24\% & 8.5\% & 2.5\% & 1.1\%  & 6.0\% \\


\end{tabular}
\label{tab:Cerrors}
\end{table}

\begin{figure}[htbp]
\centering
\includegraphics[scale = 0.4]{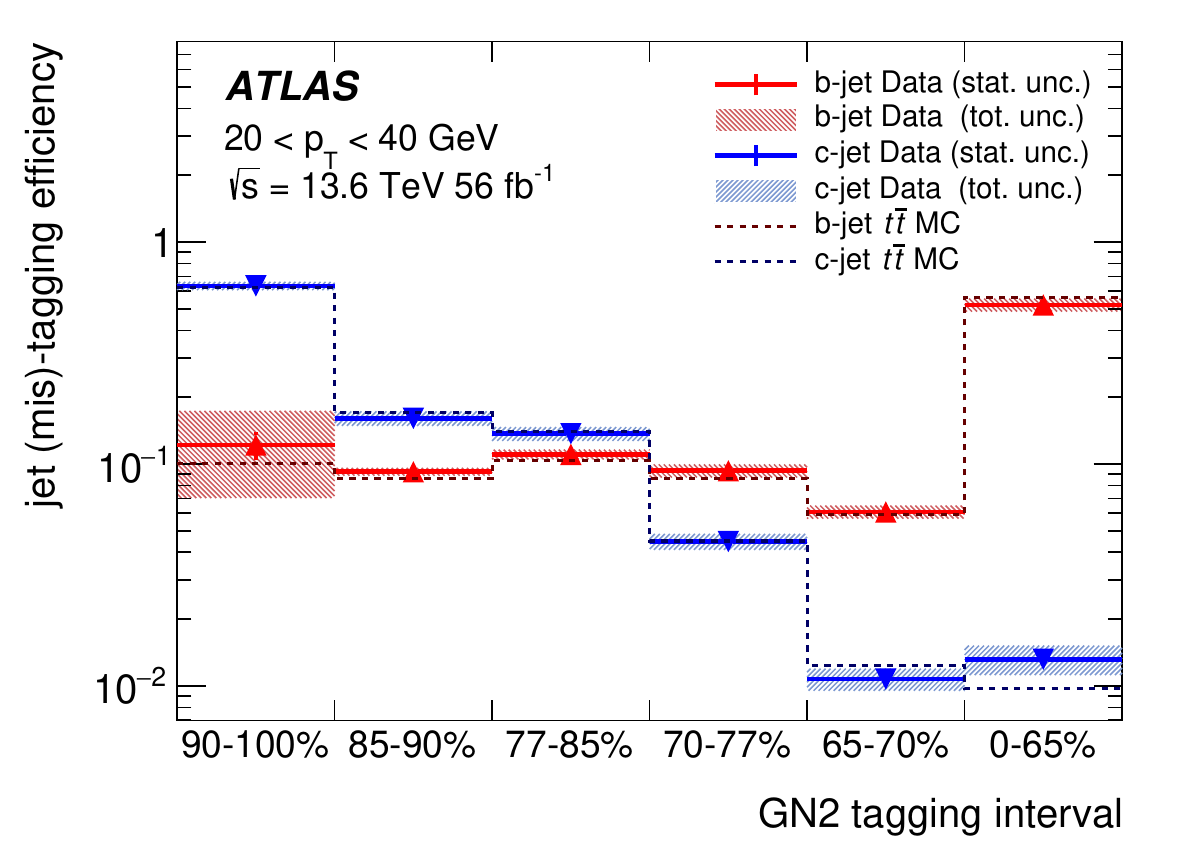}
\includegraphics[scale = 0.4]{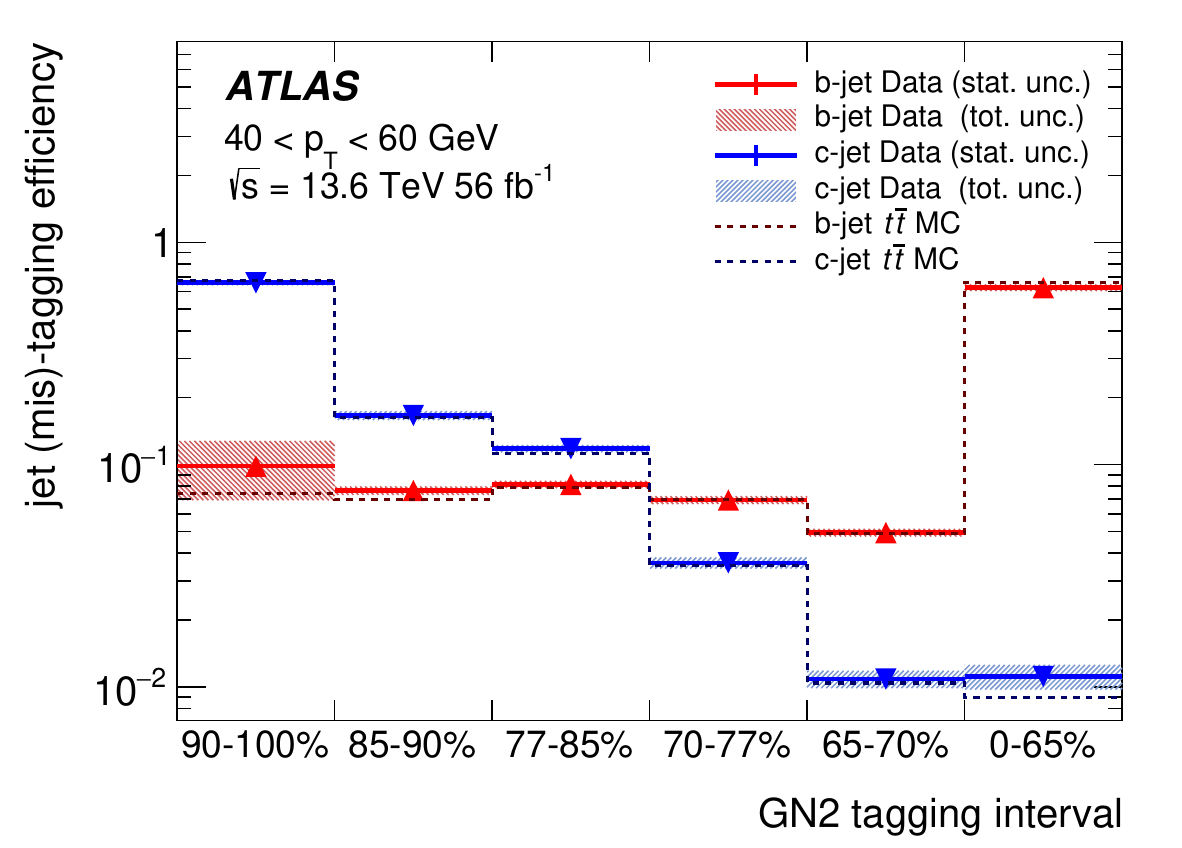} \\
\includegraphics[scale = 0.4]{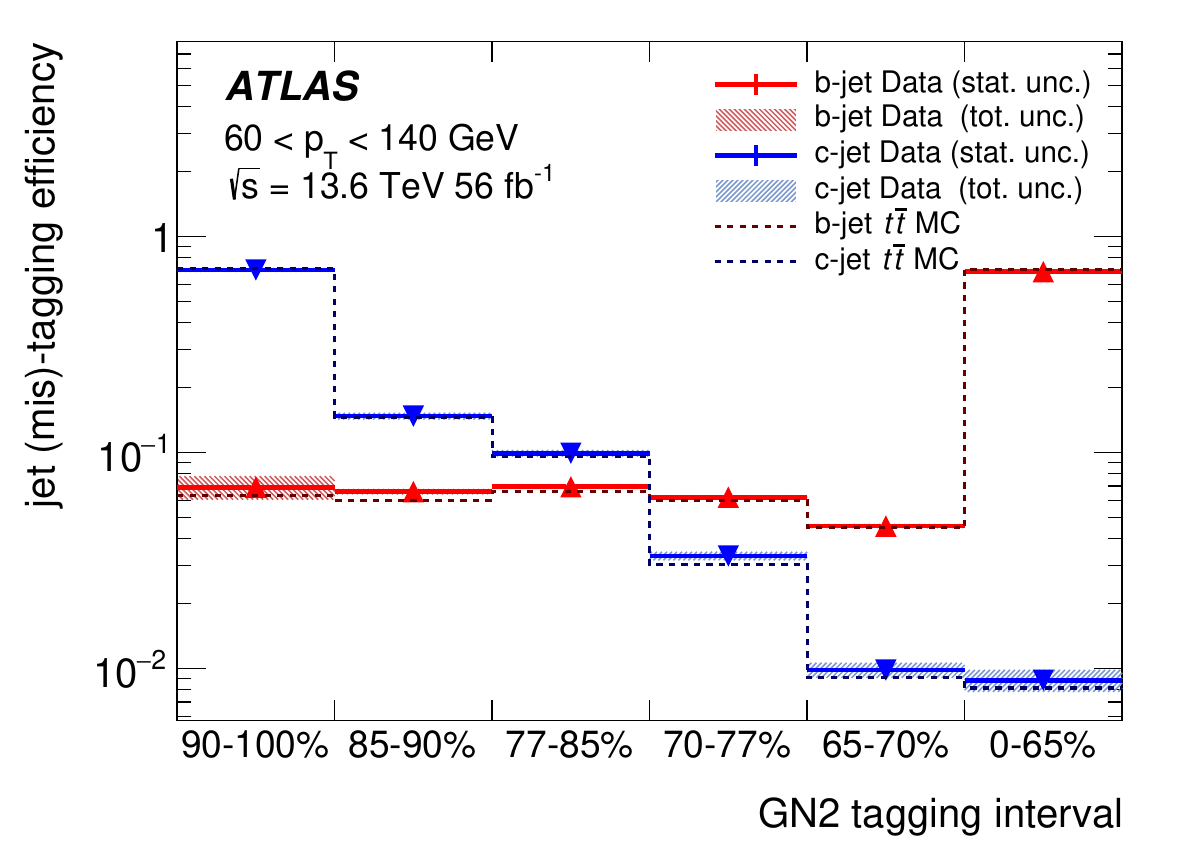}
\includegraphics[scale = 0.4]{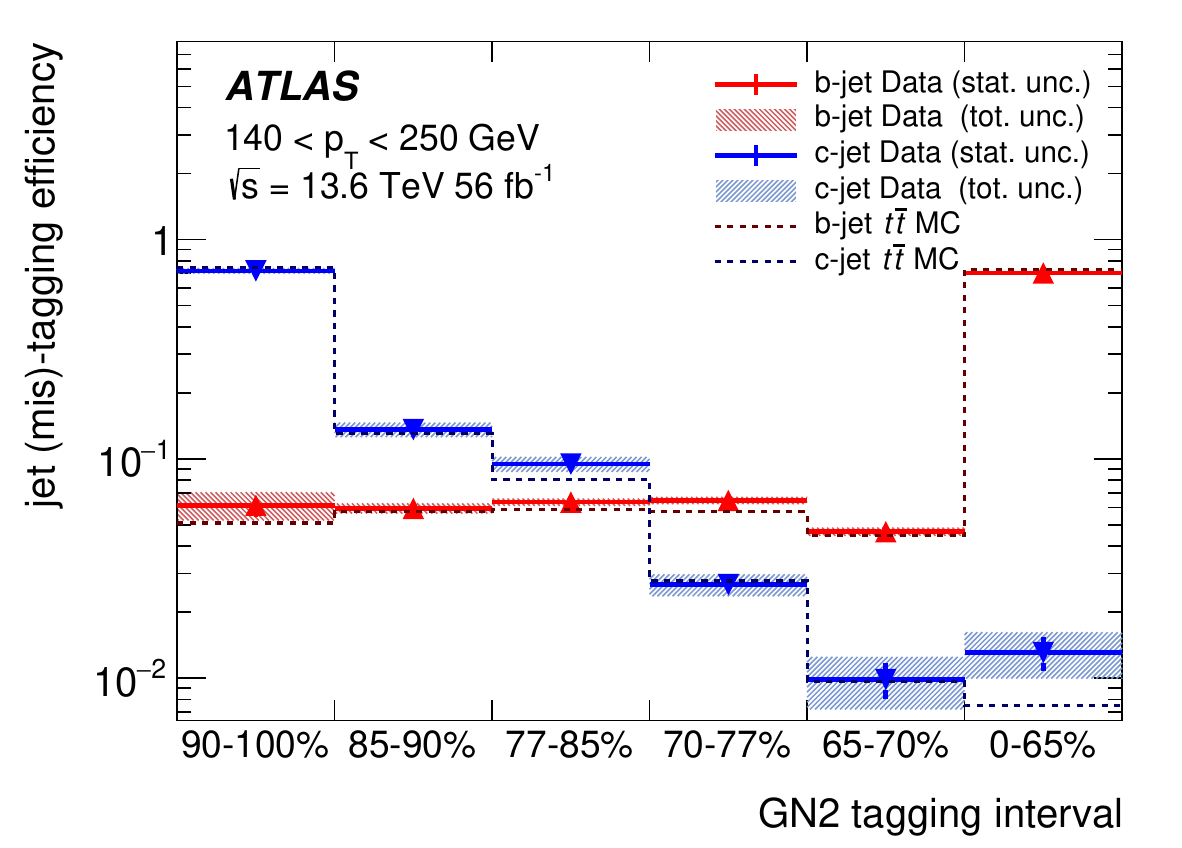}
\caption{  The \bjet\ (\cjet) pseudo-continuous (mis)tagging efficiencies
measured in  semileptonic $t\bar{t}$ events shown as points as a function of tagging interval
for the four jet-\pt\ ranges.
Included in the plots are the corresponding MC efficiencies when using the
$\PYTHIA[8]$\,+\,\EVTGEN fragmentation model.}
\label{fig:Beffs}
\end{figure}

The corresponding measured $b$-tagging efficiencies and $c$-mistagging
efficiencies, shown in Figure~\ref{fig:Beffs}, are obtained by
multiplying the tagging efficiencies from simulation by the SFs. It
can be seen that the $b$-tagging efficiency is larger than the
$c$-mistagging efficiency in the three tightest tagging intervals, whereas
the $c$-mistagging efficiency is larger in the three loosest intervals.

Post-fit plots of the GN2 discriminant are shown in Figure~\ref{fig:GNpostfit} and
Figure~\ref{fig:GNWpostfit} after applying the fitted normalisation and \cjet\ and \bjet\ SFs.
It can be seen that the fit greatly improves the description. In particular the
normalisation of the light-flavour jet background is reasonable even though
these distributions are not directly used to constrain it.
A post-fit plot of the $m_{jl}$
distribution of the charged lepton and leptonic top-jet, displayed in
Figure~\ref{fig:mblpostfit}, shows good agreement between data and
simulation, giving further confidence in the normalisation of the light-flavour jet background.

%
%
%
%
%
%

%
%
%
%
%
%
%
%
%
%
%
%

%
%
%
%
%
%
%
%
%
%
%
%
%
%
%
%
%
%
%
%
%
%
%
%
%
%
%
%
%


%
%
%

%

%
%
\FloatBarrier

%
%

%
\section{Conclusion}
\label{sec:conclusion}
Simultaneous measurements of the \bjet\  tagging and  \cjet\  mistagging efficiencies of the ATLAS GN2 \btagging algorithm are presented for the first time.
The measurements use semileptonic $t\bar{t}$ events selected from \lumi\ of $\sqrt{s}=13.6$~\TeV\ \(pp\) collision data collected with the ATLAS detector at the LHC during 2022 and 2023.
The efficiencies were measured as a
function of jet \pt\ in the range 20--250~GeV, with
boundaries corresponding to 65\%, 70\%, 77\%, 85\% and 90\% efficiency to tag \bjets\
in simulated \ttbar events.
The measured \bjet\ tagging efficiencies have a total uncertainty of
less than 2\% for jet $\pt>60$~GeV in the tightest tagging interval.
Several \cjet\ tagging efficiency measurements have a precision of better
than 5\%. These are found in various jet $\pt$ intervals in the four loosest tagging intervals.

Good agreement is observed between the measured \bjet\ efficiencies
and those obtained with an independent measurement using dileptonic
$t\bar{t}$ events. A combination of the two measurements is also
presented.


%
%
%
%
%
\section*{Acknowledgements}
%

%

%
%

%
%

We thank CERN for the very successful operation of the LHC and its injectors, as well as the support staff at
CERN and at our institutions worldwide without whom ATLAS could not be operated efficiently.

The crucial computing support from all WLCG partners is acknowledged gratefully, in particular from CERN, the ATLAS Tier-1 facilities at TRIUMF/SFU (Canada), NDGF (Denmark, Norway, Sweden), CC-IN2P3 (France), KIT/GridKA (Germany), INFN-CNAF (Italy), NL-T1 (Netherlands), PIC (Spain), RAL (UK) and BNL (USA), the Tier-2 facilities worldwide and large non-WLCG resource providers. Major contributors of computing resources are listed in Ref.~\cite{ATL-SOFT-PUB-2026-001}.

We gratefully acknowledge the support of ANPCyT, Argentina; YerPhI, Armenia; ARC, Australia; BMWFW and FWF, Austria; ANAS, Azerbaijan; CNPq and FAPESP, Brazil; NSERC, NRC and CFI, Canada; CERN; ANID, Chile; CAS, MOST and NSFC, China; Minciencias, Colombia; MEYS CR, Czech Republic; DNRF and DNSRC, Denmark; IN2P3-CNRS and CEA-DRF/IRFU, France; SRNSFG, Georgia; BMFTR, HGF and MPG, Germany; GSRI, Greece; RGC and Hong Kong SAR, China; ICHEP and Academy of Sciences and Humanities, Israel; INFN, Italy; MEXT and JSPS, Japan; CNRST, Morocco; NWO, Netherlands; RCN, Norway; MNiSW, Poland; FCT, Portugal; MNE/IFA, Romania; MSTDI, Serbia; MSSR, Slovakia; ARIS and MVZI, Slovenia; DSI/NRF, South Africa; MICIU/AEI, Spain; SRC and Wallenberg Foundation, Sweden; SERI, SNSF and Cantons of Bern and Geneva, Switzerland; NSTC, Taipei; TENMAK, T\"urkiye; STFC/UKRI, United Kingdom; DOE and NSF, United States of America.

Individual groups and members have received support from BCKDF, CANARIE, CRC and DRAC, Canada; CERN-CZ, FORTE and PRIMUS, Czech Republic; COST, ERC, ERDF, Horizon 2020 and Marie Sk{\l}odowska-Curie Actions, European Union; Investissements d'Avenir Labex, Investissements d'Avenir Idex and ANR, France; DFG and AvH Foundation, Germany; Herakleitos, Thales and Aristeia programmes co-financed by EU-ESF and the Greek NSRF, Greece; BSF-NSF and MINERVA, Israel; NCN and NAWA, Poland; La Caixa Banking Foundation, CERCA and AGAUR programs from Generalitat de Catalunya and PROMETEO and GenT Programmes Generalitat Valenciana, Spain; G\"{o}ran Gustafssons Stiftelse, Sweden; The Royal Society and Leverhulme Trust, United Kingdom; Eric and Wendy Schmidt Fund for Strategic Innovation, United States of America.

In addition, individual members wish to acknowledge support from Chile: Agencia Nacional de Investigaci\'on y Desarrollo (ANID FONDECYT reg. 1230987, FONDECYT 1230812, FONDECYT 1240864, Fondecyt 3240661, Fondecyt Regular 1240721); China: Chinese Ministry of Science and Technology (MOST-2023YFA1605700, MOST-2023YFA1609300), National Natural Science Foundation of China (NSFC - 12175119, NSFC 12275265); Czech Republic: Czech Science Foundation (GACR - 24-11373S), Ministry of Education Youth and Sports (ERC-CZ-LL2327, FORTE CZ.02.01.01/00/22\_008/0004632), PRIMUS Research Programme (PRIMUS/21/SCI/017); EU: H2020 European Research Council (ERC - 101002463); European Union: European Research Council (BARD No. 101116429, ERC - 948254, ERC 101089007), European Regional Development Fund (HE COFUND GA No.101081355, ERDF), European Union, Future Artificial Intelligence Research (FAIR-NextGenerationEU PE00000013), Marie Sklodowska-Curie Actions (GAP-101168829); France: Agence Nationale de la Recherche (ANR-21-CE31-0013, ANR-21-CE31-0022, ANR-22-EDIR-0002, ANR-24-CE31-0504-01); Germany: Deutsche Forschungsgemeinschaft (DFG - 469666862); China: Research Grants Council (GRF); Italy: Ministero dell'Università e della Ricerca (NextGenEU 153D23001490006 M4C2.1.1, NextGenEU I53D23000820006 M4C2.1.1, NextGenEU I53D23001490006 M4C2.1.1, SOE2024\_0000023); Japan: Japan Society for the Promotion of Science (JSPS KAKENHI  JP25H0063, JSPS KAKENHI JP22H01227, JSPS KAKENHI JP22H04944, JSPS KAKENHI JP22KK0227, JSPS KAKENHI JP24K23939, JSPS KAKENHI JP24KK0251, JSPS KAKENHI JP25H00650, JSPS KAKENHI JP25H01291, JSPS KAKENHI JP25K01011, JSPS KAKENHI JP25K01023); Norway: Research Council of Norway (RCN-314472); Poland: Ministry of Science and Higher Education (IDUB AGH, POB8, D4 no 9722), Polish National Science Centre (NCN 2021/42/E/ST2/00350, NCN OPUS 2023/51/B/ST2/02507, NCN OPUS nr 2022/47/B/ST2/03059, NCN UMO-2019/34/E/ST2/00393, UMO-2022/47/O/ST2/00148, UMO-2023/49/B/ST2/04085, UMO-2023/51/B/ST2/00920, UMO-2024/53/N/ST2/00869); Spain: Agència de Gestió d'Ajuts Universitaris i de Recerca. (AGAUR - 2023 BP 00141), Generalitat Valenciana (ASFAE/2022/008), Ministry of Science and Innovation (RYC2019-028510-I, RYC2020-030254-I, RYC2021-031273-I, RYC2022-038164-I), Ministerio de Ciencia, Innovación y Universidades/Agencia Estatal de Investigaci\'on (EU NextGenerationEU (PRTR-C17.I1), PID2022-142604OB-C22); Sweden: Carl Trygger Foundation (Carl Trygger Foundation CTS 22:2312), Swedish Research Council (Swedish Research Council 2023-04654, VR 2021-03651, VR 2022-03845, VR 2022-04683, VR 2023-03403, VR 2024-05451), Knut and Alice Wallenberg Foundation (KAW 2018.0458, KAW 2023.0366); Switzerland: Swiss National Science Foundation (SNSF - PCEFP2\_194658); United Kingdom: The Binks Trust, Royal Society (NIF-R1-231091); United States of America: U.S. Department of Energy (ECA DE-AC02-76SF00515), John Templeton Foundation (John Templeton Foundation 63206), Neubauer Family Foundation.

%
%


%
\clearpage
\printbibliography
\clearpage
 
\begin{flushleft}
\hypersetup{urlcolor=black}
{\Large The ATLAS Collaboration}

\bigskip

\AtlasOrcid[0000-0002-6665-4934]{G.~Aad}$^\textrm{\scriptsize 103}$,
\AtlasOrcid[0000-0001-7616-1554]{E.~Aakvaag}$^\textrm{\scriptsize 17}$,
\AtlasOrcid[0000-0002-5888-2734]{B.~Abbott}$^\textrm{\scriptsize 122}$,
\AtlasOrcid[0000-0002-0287-5869]{S.~Abdelhameed}$^\textrm{\scriptsize 84b}$,
\AtlasOrcid[0000-0002-1002-1652]{K.~Abeling}$^\textrm{\scriptsize 55}$,
\AtlasOrcid[0000-0001-5763-2760]{N.J.~Abicht}$^\textrm{\scriptsize 49}$,
\AtlasOrcid[0000-0002-8496-9294]{S.H.~Abidi}$^\textrm{\scriptsize 30}$,
\AtlasOrcid[0009-0003-6578-220X]{M.~Aboelela}$^\textrm{\scriptsize 45}$,
\AtlasOrcid[0000-0002-9987-2292]{A.~Aboulhorma}$^\textrm{\scriptsize 36e}$,
\AtlasOrcid[0000-0001-5329-6640]{H.~Abramowicz}$^\textrm{\scriptsize 155}$
% ...
\bigskip
\\

% ...
$^{17}$Department for Physics and Technology, University of Bergen, Bergen; Norway.\\
% ...
$^{30}$Physics Department, Brookhaven National Laboratory, Upton NY; United States of America.\\
% ...
$^{36}$$^{(a)}$Facult\'e des Sciences Ain Chock, Universit\'e Hassan II de Casablanca;$^{(b)}$Facult\'{e} des Sciences, Universit\'{e} Ibn-Tofail, K\'{e}nitra;$^{(c)}$Facult\'e des Sciences Semlalia, Universit\'e Cadi Ayyad, LPHEA-Marrakech;$^{(d)}$LPMR, Facult\'e des Sciences, Universit\'e Mohamed Premier, Oujda;$^{(e)}$Facult\'e des sciences, Universit\'e Mohammed V, Rabat;$^{(f)}$Institute of Applied Physics, Mohammed VI Polytechnic University, Ben Guerir; Morocco.\\
% ...
$^{45}$Physics Department, Southern Methodist University, Dallas TX; United States of America.\\
% ...
$^{49}$Fakult\"{a}t Physik, Technische Universit{\"a}t Dortmund, Dortmund; Germany.\\
% ...
$^{55}$II. Physikalisches Institut, Georg-August-Universit\"{a}t G\"ottingen, G\"ottingen; Germany.\\
% ...
$^{84}$$^{(a)}$Khalifa University of Science and Technology, Abu Dhabi;$^{(b)}$New York University Abu Dhabi, Abu Dhabi;$^{(c)}$United Arab Emirates University, Al Ain;$^{(d)}$University of Sharjah, Sharjah; United Arab Emirates.\\
% ...
$^{103}$CPPM, Aix-Marseille Universit\'e, CNRS/IN2P3, Marseille; France.\\
% ...
$^{122}$Homer L. Dodge Department of Physics and Astronomy, University of Oklahoma, Norman OK; United States of America.\\
% ...
$^{155}$Raymond and Beverly Sackler School of Physics and Astronomy, Tel Aviv University, Tel Aviv; Israel.\\
% ...
$^{a}$ Also at Affiliated with an institute formerly covered by a cooperation agreement with CERN.\\
$^{b}$ Also at An-Najah National University, Nablus; Palestine.\\
$^{c}$ Also at Borough of Manhattan Community College, City University of New York, New York NY; United States of America.\\
$^{d}$ Also at Center for Interdisciplinary Research and Innovation (CIRI-AUTH), Thessaloniki; Greece.\\
$^{e}$ Also at Centre of Physics of the Universities of Minho and Porto (CF-UM-UP); Portugal.\\
$^{f}$ Also at CERN, Geneva; Switzerland.\\
% ...

\end{flushleft}


%

%
%
%
%

%
%
%
%

\end{document}